\documentclass[
]{elteikthesis}[2023/04/10]

\usepackage{color,soul}
\usepackage{algorithm}
\usepackage{algpseudocode}
\usepackage{booktabs}
\usepackage{dsfont}
\usepackage{amsmath} 
\usepackage{amssymb} 
\usepackage{tikz,lipsum,lmodern}
\usepackage[most]{tcolorbox}

\setulcolor{red}
\title{A Gentle Introduction to \\ Blind signatures: From RSA to Lattice-based Cryptography} 
\date{2024} 

\author{Bhardwaj Aditya}
\degree{Computer Science MSc Cybersecurity}

\supervisor{Dr. Péter Kutas} 
\affiliation{Associate Professor} 

\university{Eötvös Loránd University} 
\faculty{Faculty of Informatics} 
\department{Dept. of Software Technology and Methodology} 
\city{Budapest} 
\logo{elte.png} 

\begin{document}

\documentlang{english}


\maketitle

\begin{abstract}

Blind signatures were first introduced by David Chaum. They allow a user to have a message signed by a signer without revealing the message itself. This property is particularly useful in applications such as electronic voting and digital cash, where user anonymity is important. In a blind signature scheme, the user blinds their message before sending it to the signer, who signs the blinded message. The user then unblinds the signed message to obtain a valid signature that can be verified publicly, ensuring that the signer cannot trace the signed message back to the original unblinded version. A good analogy is placing the message inside an envelope and having the envelope signed. Once the envelope is opened, the signature remains valid for the enclosed message, ensuring that the content remains confidential. 

Such constructions provide anonymity and privacy to the user but given a practical quantum computer, the security of traditional cryptosystems providing such features will be broken. To address this, the development of quantum-resistant cryptographic protocols is essential for maintaining the security of digital transactions and data. Aligning with the same goal, this work aims to thoroughly review the background of lattice-based blind signatures and then look at an application - a scheme for $ecash$ or digital cash, which we can adopt for providing post-quantum security.
\end{abstract}

\tableofcontents
\cleardoublepage

\chapter*{\acklabel}
\addcontentsline{toc}{chapter}{\acklabel}
I learnt a lot of things while I was writing the thesis. I would like to thank my supervisor Dr. Péter Kutas who guided me along the way and cleared all my doubts. My parents were very supportive of me and made sure I could focus solely on my studies. Lastly I would like to thank ELTE for giving me the opportunity to study and last but not the least... my friends, who always stayed by my side.

\chapter{Introduction}
\label{ch:intro}

Public-key cryptography which is another name for asymmetric cryptography, is at the heart of current communication systems. In symmetric cryptography, both sides - sender and receiver, share the same secret key. In public-key cryptography, on the other hand, there are two keys, one public and one private. This allows anyone to encrypt messages with the public key, but only the holder of matching private key can decrypt them. Many security protocols are based on this method, which makes sure that communication over unprotected channels like the Internet is safe, confidential and authenticated. Transport Layer Security (TLS), Secure Shell protocol (SSH), IPSec, Pretty Good Privacy (PGP) are some of the well-known protocols that employ public-key encryption in their schemes. 

In the year 1983, public-key cryptography was still in its early days when David Chaum devised eCash \cite{chaumblind} in his paper titled \textit{Blind Signatures for Untraceable Payments}. The goal was to create an electronic form of cash that would have the same privacy and secrecy as real cash. This notion was revolutionary, particularly during this time of the internet when public-key cryptography was being developed. To realize this idea, he introduced \textit{blind} signatures, a form of digital signature in which the content of a message is blinded or hidden from the signer before it is signed. Hence, in the context of eCash, a masked digital cash token is made during token generation and is then signed by a bank. While the blind signature confirms that the token is authentic and is indeed signed by the bank, the signer doesn't see the message content. The person then 
$unblinds$ or unmasks the token, which shows the original digital token that was signed. This process protects privacy and security by enabling transactions without being able to link the person to the original token. In e-voting systems \cite{evoting}, blind signatures are used to maintain voter privacy by allowing ballots to be signed without revealing the voter's choices. In credential systems \cite{baldimtsi2013anonymous}, it helps users prove credentials without revealing themselves, addressing privacy concerns in access control systems.

The basis of security in public key cryptography today largely relies on the computational difficulty of certain mathematical problems like the \textit{Integer Factorization Problem} or \textit{Discrete Logarithm Problem}. These problems are already broken theoretically by Shor algorithm \cite{shor1994algorithms} but it assumes a quantum computer free of noise and error, which is still largely not practical. Given the current computing technology that includes both classical and current quantum computing capabilities, it still remains infeasible to solve these problems (for the relevant parameters) within a reasonable frame of time.  One can still not downplay the threat which is posed by the continuous advancements in quantum computing. It is necessary to create new cryptographic algorithms and modify existing ones to maintain continuous security. Post-quantum cryptography is a field that deals with cryptographic problems that are considered to be resistant to attacks from quantum computers. 

In 2016, NIST \cite{nist2016submission} started a process to assess, and standardize a few public-key cryptography algorithms that are based on such problems. The objective was to standardise algorithms that are resistant to attacks from both classical and quantum computers. In 2022, four algorithms were selected for the standardization - CRYSTALS-Kyber \cite{bos2018crystalskyber}, CRYSTALS-Dilithium \cite{ducas2018crystalsdilithium}, SPHINCS+ \cite{bernstein2019sphincs+} and FALCON \cite{prest2020falcon}. While CRYSTALS-Kyber is a key encapsulation mechanism, the others are digital signature schemes. Out of these four algorithms, three are based on lattice-based cryptography which is the focus of this work. Lattice structure are discrete sets of points in multidimensional space formed by linear combinations of vectors called a \textit{basis}. For the majority of real-world applications, lattice-based cryptography's overall performance is adequate and can develop powerful tools for key exchange, homomorphic encryption, attribute-based encryption, signatures and hash functions. 

\section{Objective}

There are different approaches to post-quantum cryptography and these are Lattice-based, code-based, multivariate-based, hash-based, isogeny-based, and hash-based schemes. Their security is based on different mathematical problems that are believed to be resistant to quantum attacks. This work focuses on lattice-based cryptography which is one of the most potential PQC candidates due to its advantages \cite{wang2023lattice}: 

\begin{itemize}
    \item Strong security: Ajtai \cite{ajtai1996generating} gave the worst-case to average-case reduction for lattice problems. It states that a `randomly chosen instance of a certain lattice problem is at least as hard as the worst-case instance of a related lattice problem'. The two hard problems in lattice, Shortest Vector Problem SVP and CVP are shown to be NP-hard and moreover, the underlying structure of lattice-cryptography work in high dimensions, which can be several hundreds or thousands.
    \item Performance: Lattice-based schemes are on par with the current public key algorithms. Even though lattice-based cryptography creates larger keys than of RSA and ECC schemes, it has a balanced performance and efficiency compared to other post-quantum cryptosystems. And works \cite{gentry2009fully}, \cite{GPV08} by researchers like Gentry, Vaikuntanathan, Lyubashevsky, Chris Peikert have provided us efficient lattice-based constructions for a wide variety of applications.

\end{itemize}
Since Ruckert et al. \cite{ruckert2010lattice} gave the first lattice-based blind signature in 2010, constructions have also been given for specific scenarios including group, ring, blind, and proxy signatures. The main topic of this work is lattice-based blind signatures and the objectives are as follows:

\begin{itemize}
   \item Give an overview of lattice-based cryptography and the main construction tools available.
   \item Review the recent lattice-based blind signature schemes and the new security assumptions they are based on.
   \item Study some recent attacks and list the schemes affected by these attacks
   \item Describe the construction of the first blind signature scheme to have a formal security proof in the quantum random oracle model
   \item Review the approaches that tackle the problem of double-spending in electronic cash and briefly evaluate their feasibility if lattice-based blind signatures are used.
\end{itemize}

\section{Outline}

We briefly discussed about public-key cryptography and then blind signatures, where the signer signs on a message without seeing the message itself. The underlying security of today's digital infrastructure relies on the fact that some mathematical problems are hard to solve (in a feasible amount of time). A practical and strong enough quantum computer could pose a threat to the same and hence development of post-quantum cryptography is evident.  

The goal is to familiarize the reader with blind signatures and then discuss the construction of a lattice-based blind signature scheme finally. We talk briefly about digital signatures in Chapter \ref{ch:chapter_2}. In Chapter \ref{ch:chapter_3}, we discuss blind signatures in a little detail along with $ROS$ problem. Lattice-based cryptography is one of the techniques that use the mathematical structure of lattices. Some problems in lattices are proven to be resistant against quantum attacks. We discuss these and the construction of lattice-based signatures in Chapter \ref{ch:chapter_4}. Finally we discuss the state of the art in lattice-based blind signatures in Chapter \ref{ch:chapter_5} and understand the construction of the scheme given by del Pino and Katsumata et al. \cite{del2022new}. 

\cleardoublepage

\chapter{Digital Signatures}
\label{ch:chapter_2}

Until the mid-1970s, all cipher systems used symmetric cryptography. In this system, a single key is used to encrypt and decrypt the data. The private or secret key must be known to both sender and receiver. While this is fast and efficient, for $n$ parties, this involves $n(n-1)$ keys to be distributed. If an eavesdropper intercepts the key using transmission, they can decrypt and see the messages which brings us to the issue of authentication and trust between the communicating parties. In 1976, Whitfield Diffie and Martin Hellman introduced the concept of public key cryptography in their landmark paper ``New Directions in Cryptography,'' \cite{diffie1976newdirections}. (they were influenced by Merkle’s work \cite{merkle1978} who discussed about secure transmission of keys in insecure channel but was not practical). They proposed the idea of a private key and a corresponding public key which allows the parties to establish a shared secret-key over insecure channel; this is known as the Diffie-Hellman key exchange. Along with asymmetric key cryptography, Diffie and Hellman were also the first one to give the notion of digital signature scheme.

Like paper-based signatures, digital signatures could provide two guarantees: $Authentication$, that a message came from the right person and  $Non-repudiation$, that a signer of a message cannot deny and claim they did not sign it. Diffie and Hellman had conjectured that such schemes use one-way functions but it was still unknown whether such functions exist. It was in 1978 when Rivest, Shamir and Adleman gave the construction of first digital signature scheme which was based on the RSA public-key cryptosystem \cite{rsapaper}. It’s security lies in the extreme difficulty of factoring large integers and the modular exponentiation function works as the one-way function. 

\section{Applications}

A digital signature algorithm enables an entity to verify both the integrity of signed data and the identity of the signer. When receiving a signed message, the recipient can present the digital signature as proof to a third party that it indeed originated from the intended signer. This concept, known as non-repudiation, prevents the signer from easily disowning the signature afterward. Digital signature algorithms find application in various domains, including electronic mail, electronic funds transfer, electronic data interchange, software distribution, data storage, and other contexts where ensuring data integrity and authenticating data origin are essential. In a practical setting, digital certificates are used in Public Key Infrastructure (PKI) which helps establishes trustworthiness among the entities (users, devices and networks). Digital certificate act as a proof of ownership of a public key which also contains other information about the entity along with digital signature. PKI sets policies, roles and creates, distributes and can revoke (if needed) these digital certificates. It is the underlying infrastructure which allows safe communication (TLS protocol), secure e-mail (PGP) or digital signing of applications or software.

NIST publishes Federal Information Processing Standard (FIPS) in which they also share the schemes that can be used for generation and verification of digital signature so as to promote secure, interoperable, and standardized technologies. The three algorithms are Rivest-Shamir-Adleman (RSA) Algorithm, Elliptic Curve Digital Signature Algorithm (ECDSA) and Edwards Curve Digital Signature Algorithm (EdDSA) \cite{NISTDigSign}.

\section{Digital signature scheme}

A digital signature scheme typically consists of three algorithms ($KeyGen$, $Sign$, $Verify$), they are as follows:

\begin{description}
	\item [\textit{\textbf{KeyGen($k$)}}] With an input parameter $k$, this algorithm generates a private key which is selected from a set of possible private keys. We get the pair of cryptographic keys ($sk$,$pk$) where $sk$ is the private key and $pk$ is the corresponding public key pk. We can denote it as KeyGen($k$) $\rightarrow$ ($sk$, $pk$)

    \item [\textbf{\textit{Sign($m,sk$)}}] Given the message $m$ and private key $sk$ as the input, signing algorithm generates the signature $\pi$. This is denoted as Sign($m$,$sk$) $\rightarrow$ $\pi$

    \item [\textbf{\textit{Verify($m,$$\pi,pk$)}}] Finally, this deterministic  algorithm takes as input a message $m$, the signature $\pi$ to verify and the public key $pk$. It returns true or bit $1$ if ($m$, $\pi$) is a valid message-signature pair. Otherwise, the signature is rejected and algorithm returns false or 0 bit. We can denote it as Verify($m$,$\pi$,$pk$)  $\rightarrow$ ($true$/$false$).
 
\end{description}

\section{Security of Digital Signatures}

Unforgeability is a fundamental security property of digital signature schemes, ensuring that for a public key $p k$, only the legitimate signer $S$ can create valid signatures $\sigma$ for a given message $m$. It guarantees that no adversary can forge a signature on a new message even with knowledge of other signed messages (given by the signer $S$). For the formal definition of security, we refer to \cite{katz2007introduction}. Let the signature scheme be $\Pi=($ Gen, Sign, Vrfy). For an adversary $\mathcal{A}$ and given parameter $n$, the signature experiment Sig-forge ${ }_{\mathcal{A}, \Pi}(n)$ is as follows:

1. $\operatorname{Gen}\left(1^n\right)$ $\rightarrow$ key pair $(p k, s k)$ .

2. Given $p k$ and access to an oracle $\operatorname{Sign}_{s k}(\cdot)$, the goal of the adversary $\mathcal{A}$ is to generate the message-signature pair $(m, \sigma)$ for the message $m$ which he has not queried to the signing oracle $\operatorname{Sign}_{s k}(\cdot)$ before. Let $\mathcal{Q}$ represent all queries that $\mathcal{A}$ sent to the oracle.

3. $\mathcal{A}$ wins this experiment if $m \notin \mathcal{Q}$ and $\operatorname{Vrfy}_{p k}(m, \sigma)=1$ i.e. $(m, \sigma)$ is a valid pair. In this case the output of the experiment is defined to be 1 and otherwise 0. 

We summarize this discussion in the following definition (\cite[Definition 12.2]{katz2007introduction}):

\begin{definition}
A signature scheme $\Pi=($ Gen, Sign, Vrfy) is existentially unforgeable under an adaptive chosen-message attack, or just secure, if for all probabilistic polynomial-time adversaries $\mathcal{A}$, there is a negligible function $\mathrm{negl}$ such that:
$$
\operatorname{Pr}\left[\text { Sig-forge }_{\mathcal{A}, \Pi}(n)=1\right] \leq \operatorname{\mathrm{negl}}(n) .
$$
    
\end{definition}

\section{Hashing}

In digital signatures, hashing is used to convert long-arbitrary message input into a fixed length value called hash.  In digital signature schemes, we sign the hash value rather than the message itself. The entity that receives the message and the corresponding signature, first derives the message's hash and then confirms if the provided signature is valid for the hash value. 
 
 \begin{definition}
     For an arbitrary input, a hash function is a function H that gives us a fixed-length output. This is simply called the \textit{hash} or the message \textit{digest}. We can denote the hash function H as: 
    \[ y = H:\{0, 1\}^* \rightarrow \{0, 1\}^k \] where $k$ denotes the length of the output.

Along with the above property of input-compression, it is desirable to have the following properties:

\begin{enumerate}
    \item Fast and Ease of computation: It should be easy to compute the hash $H(x)$ for an input $x$.
    \item Preimage resistance: Given a hash value $h$, it should be computationally infeasible to find any input $x$ such that $H(x) = h$.
    \item Second Pre-image resistance: For any given input $x_1$, it should be computationally infeasible to find another input $x_2$ such that $H(x_1) = H(x_2)$.
    \item Collision Resistance: It should be computationally infeasible to find two inputs $x_1$, $x_2$ where $x_1$ $\neq$ $x_2$ and $H(x_1) = H(x_2)$.
\end{enumerate}
 \end{definition}

\section{Computational Hard Problems}

There are computational tasks for which we (yet) do not have efficient algorithms to solve them in polynomial time. Number theory presents us these problems \cite{Gaudry2014IntegerFA} where the amount of computational resources (such as time and memory) required to solve them grows exponentially with the size of the input. The most widely used problems which are used in the construction of public-key cryptosystems and hence signature schemes are as follows:

\begin{enumerate}
    \item Integer Factorization Problem (IFP): This problem is the basis of the popular algorithms like RSA \cite{rsapaper} and Rabin \cite{RabinIFP1978}. It is as following: given a positive integer n, find the prime factorization with the components $e_i$ and $p_i$ such as
    \[ N = {p_1}^{e_1} {p_2}^{e_2} ... {p_k}^{e_k}\] where $p_i$ are the prime numbers and $e_i$ $\geq$ 1
    This problem is computationally difficult for large numbers with hundreds or thousands of digits.
    
    \item The Discrete Logarithm Problem (DLP): It involves finding an integer $x$ such that $g^x=h$, where $g$ is a generator of a finite cyclic group $G$ of order $p$, and $h$ is an element in $G$. The security of several cryptographic protocols, such as the Diffie-Hellman (DH) key exchange, Digital Signature Algorithm (DSA), and ElGamal encryption relies on the assumption that the DLP is computationally difficult to solve.
    
    \item Elliptic Curve Discrete Logarithmic Problem (ECDLP): Given an elliptic curve point $P$, a generator point $G$, and a prime $p$, find $k$ such that $P = k \cdot G$ in the elliptic curve group over $\mathbb{F}_p$. The integer $k$, denoted as $k$ = ${log_P}Q$, is called the discrete logarithm of $P$ and is hard to find. The elliptic curve variant schemes, Elliptic Curve Diffie-Hellman (ECDH) and Elliptic Curve Digital Signature Algorithm (ECDSA) are some schemes whose security rely on this problem.
\end{enumerate}

\section{Digital Signature schemes}

Now we describe some of the most notable digital signature schemes namely RSA (named after the authors Rivest–Shamir–Adleman)and Digital Signature Algorithm (DSA) . While RSA is based on the integer factorization problem, DSA relies on the discrete logarithm problem. 

\subsection{RSA Digital Signature scheme}

\subsubsection*{Key generation}

To set up the RSA signature scheme and create signatures, Alice (the signer) needs to generate her pair of public and private key. She chooses two large prime numbers, \( p \) and \( q \), and computes the product \( n = pq \), the $modulus$.  Alice also calculates the Euler's totient \( \phi(n) = (p-1)(q-1) \). 

While the modulus $n$ is published as a part of the public key, $\phi(n)$ is kept secret. Alice selects an integer \( e \) which is co-prime to $\phi(n)$, it means \( 1 < e < \phi(n) \) and \( \gcd(e, \phi(n)) = 1 \). This integer \( e \) is also part of the public key. Alice finally computes the private exponent \( d \) such that \( d \equiv e^{-1} \mod \phi(n) \) or \( de \equiv 1 \mod \phi(n) \). This means \( d \) is the modular inverse of \( e \) modulo \( \phi(n) \).

The public key is \( (e, n) \) and the private key is \( (d, n) \).

\subsubsection*{Signing}

Given a message $m$ and her private key $(d,n)$, Alice first calculates the hash or the message digest $h \leftarrow H(m)$ where $H(m)$ is a hash function.  With the private exponent $d$ , she generates the signature $s$ : $s \leftarrow h^d$ $mod$ $n$. 

Alice can now send the message $m$ along with the signature $s$ to Bob (the verifier).

\subsubsection*{Verification}

Bob has Alice's public key $(e,n)$. Given the message-signature pair $(m,s)$, he computes the hash value $h$  of the message $m$. He computes  $s^e \mod n$ to get $h'$ i.e. $h'\leftarrow s^e \mod n$ and can check if $h = h'$. If they match, Bob concludes that the signature $s$ is valid and the message $m$ was indeed signed by Alice. 

This is because:
\begin{center}
\( h' = s^e \mod n = (h^d)^e \mod n = h\)    
\end{center}

\subsection{DSA Digital Signature scheme}

\subsubsection*{Key generation}

The key generation phase has two parts. First the parameters are selected and then the key pair is generated. Digital Signature Standard (DSS) \cite{nistDSSfips} specifies the parameter choices. A prime number $p$ is chosen such that $p-1$ is a multiple of a large prime $q$. DSS has specified the length of $p$ and $q$ as well. For example if prime $q$ is of 256-bit length, then $p$ can be a 2048-bit or a 3072-bit prime. Also, a generator $g$ of the subgroup of order $q$ in $\mathbb{Z}_p^*$ is taken.
        
To generate the private and public keys, Alice generates a random integer $x$ from $\{1,q-1\}$. This is her private key. She computes the public key as $y = g^x \mod p$.

\subsubsection*{Signing}

The signature of a message $m$ consists of the pair of numbers $r$ and $s$. 

Alice chooses a random integer $k$ from $\{1,q-1\}$ and then computes $r = (g^k \mod p) \mod q$. If $r = 0$, she chooses a different $k$. Alice calculates the hash value $h \leftarrow H(m)$ and then $s = k^{-1}(h + xr) \mod q$. She finally publishes the message $m$ and signature $(r,s)$.

\subsubsection*{Verification}
Bob obtains $(p, q, g, y)$ which contains Alice's public key and public parameters. Given the message $m$ and signature $(r,s)$, he computes the hash value $h \leftarrow H(m)$.
and $w = s^{-1} \mod q$.

Bob calculates $u_1 = h \cdot w \mod q$ and $u_2 = r \cdot w \mod q$.
He computes $v = ((g^{u_1} \cdot y^{u_2}) \mod p) \mod q$.
If $v = r$, the signature is valid; otherwise, it's invalid.

This is because
\begin{center}
$(g^{u_1} \cdot y^{u_2}) \mod p$ = $(g^{hw} \cdot y^{rw}) \mod p$

$(g^{u_1} \cdot y^{u_2}) \mod p$ = $(g^{hw} \cdot (g^x)^{rw}) \mod p$

$(g^{u_1} \cdot y^{u_2}) \mod p$ = $(g^{hw+xrw}) \mod p$
\end{center}

Just taking the exponent, the value is 

\begin{center}
$hw+xrw = w(h+xr)$
\end{center}

Earlier we saw that $w = s^{-1} \mod q$. Then

\begin{center}
$w(h+xr)$ = $s^{-1}(h + xr) \mod q$

$w(h+xr)$ = $k \cdot (h + xr)^{-1}(h + xr) \mod q$

$= k \mod q$
\end{center}

Hence $(g^{hw+xrw})\mod p$ = \( (g^k\mod p) \mod q \) = $r$

\subsection{ECDSA}

Elliptic Curve Digital Signature Algorithm (ECDSA) is the elliptic curve variant of DSA. Both signer and verifier, agree on some parameters that define the elliptic curve. NIST publishes their recommendation for such curves and parameters \cite{chen2019recommendations}.

\subsubsection*{Key generation}

To generate the key pair and send signed messages, Alice and Bob first decide on the domain parameters, these are as follows:

\begin{itemize}
    \item $p$, a large prime number
    \item coefficients $a$ and $b$ defining the elliptic curve $E$
    \item $G = (x_G, y_G)$ a base point on the curve $E$ that generates the subgroup
    \item $n$, the order of this subgroup
\end{itemize}   

Alice selects a random integer $d$ from $\{1,n-1\}$. This will be her private key. She computes the public key $Q = d \cdot G$.

\subsubsection*{Signing}
Alice first generates a random integer $k$ from $\{1,n-1\}$ and calculates the curve point $P = k \cdot G = (x_P, y_P)$. She calculates $r = x_P \mod n$ and $s = k^{-1}(h + dr) \mod n$, where $h \leftarrow H(m)$ is the hash of the message $m$ and $d$ is her private key. Alice publishes the signature pair $(r,s)$.

\subsubsection*{Verification}

Given Alice's public key $Q$, the base point $G$, the message $m$ and the signature pair $(r,s)$, Bob first computes $w = s^{-1} \mod q$ and hash value $h \leftarrow H(m)$.

Bob calculates $u_1 = hw \mod n$ and $u_2 = rw \mod n$ and computes the elliptic curve point in the following way:
\[
(x_2, y_2) = u_1 G + u_2 Q
\]

The signature is valid if $r \equiv x_2 \mod n$, otherwise it's invalid.

\subsubsection{Correctness}

We recall that $w = s^{-1} \mod q$ and Bob computed $u_1 = hw \mod n$ and $u_2 = rw \mod n$. Substituting $w$ in the equations of $u_1, u_2$:
\[
u_1 = h s^{-1} \mod n \quad \text{and} \quad u_2 = r s^{-1} \mod n
\]
\[
(x_2, y_2) = (h s^{-1}) G + (r s^{-1}) Q
\]
Since \( Q = d \cdot G \), this becomes:
\[
(x_2, y_2) = (h s^{-1}) G + (r s^{-1}) dG = (h s^{-1} + r d s^{-1}) G
\]
Take the expression inside the parenthesis:
\[
x_2 \equiv (h s^{-1} + r d s^{-1}) \mod n \equiv ((h + rd) s^{-1}) \mod n
\]
During signing step, computed :
\[
s = k^{-1}(h + dr) \mod n \Rightarrow \quad k \equiv (h + rd) s^{-1} \mod n
\]
Therefore:
\[
x_2 \equiv k \cdot G \mod n
\]
\[
x_2 \equiv r \mod n
\]

\section{Security}

\begin{center}
\begin{table}[htbp]
  \centering
    \begin{tabular}{c|c|c}
    
    Security level (bits) & RSA & Elliptic Curve - ECDSA, EdDSA \\
    \hline
    80    & 1024  & 160 \\ 
    \hline
    112   & 2048  & 224 \\
    \hline
    \textbf{128}   & \textbf{3072}  & 256 \\
    \hline
    \textbf{192}   & \textbf{7680}  & \textbf{384} \\
    \hline
    \textbf{256}   & \textbf{15360} & \textbf{521} \\
    \hline
    \end{tabular}
  \caption{Key-size (bits) for same level of security. Selections in bold are recommended by NIST}
  \label{tab:key_size}
\end{table}
\end{center}

Table \ref{tab:key_size} shows the security provided by the schemes RSA and elliptic curve based ECDSA and EdDSA. When we say that a scheme provides security level of 128 bits, it refers to the brute-force attack where the adversary would need $2^{128}$ operations to succeed. NIST guidance says that ``\textit{the use of keys that provide less than 112 bits of security strength for key agreement is now disallowed}'' \cite{barker2020nist}.

The security of RSA relies on the hardness of factoring integers. If the attacker can successfully factor the public modulus $N=pq$ available in the public key, they can compute the private key as well. Finding the prime factors $p$ and $q$ appears to be the logical approach in breaking RSA but there might be ways of decrypting RSA ciphertexts or forging RSA signatures without factorization. Hence we know that this RSA problem is as easy as factoring and not vice-versa.

\begin{figure}
    \centering
    \includegraphics[width=12cm, height=8cm] {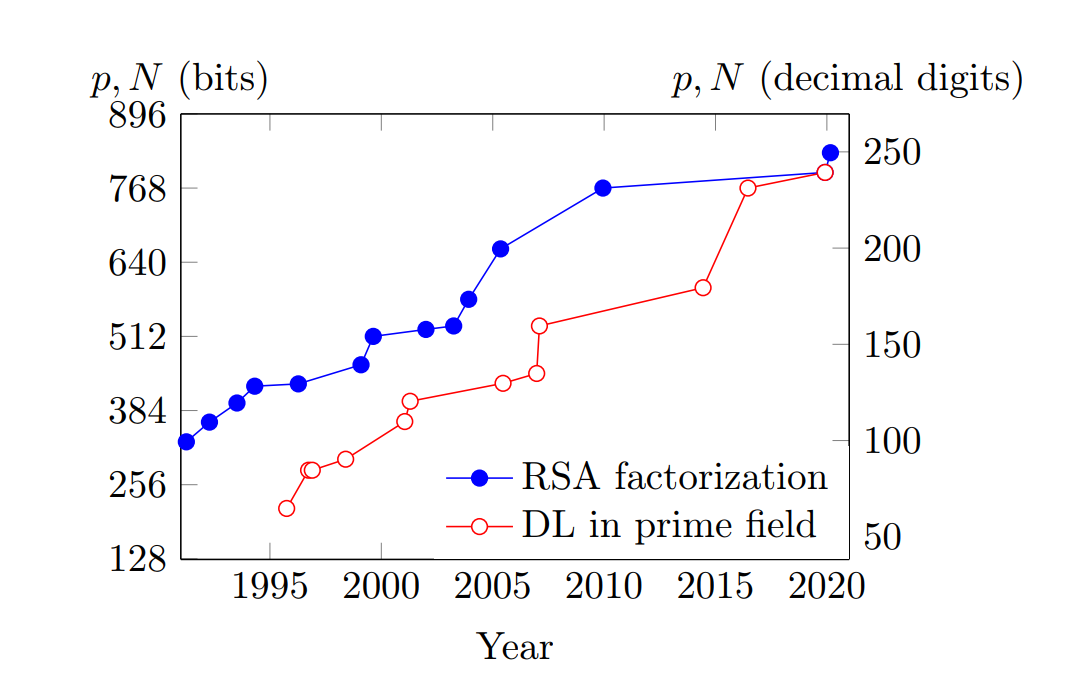}
    \caption{Current RSA and DL breaking records \cite{boudot2022state}}
    \label{fig:rsa-dl-records}
\end{figure}

In practice, cryptographic algorithms use \textit{semiprime} (product of two prime numbers), whose factors are similar in size. These are the most difficult to factor. Figure \ref{fig:rsa-dl-records} shows the current factoring and discrete logarithm records. In 2020, \cite{rsa250break} showed the factoring of RSA-250, a 829-bit number and before this, \cite{RSAbreak2020} completed factorization of RSA-240,
a 795-bit number, and a discrete logarithm computation over a 795-bit
prime field. The authors used the \textit{Number Field Sieve} algorithm to achieve the same in both the cases.

There are classical algorithms to factor numbers if their prime factors have \textit{certain} forms. For example Pollard $p-1$ algorithm can find a factor $p$ of an integer if the $p-1$ value has small prime factors. Some other algorithms are Pollard's \textit{rho} and Fermat's factorization \cite{boudot2022state}. These types of algorithms are applied before general purpose algorithms because cryptographic constructions avoid usage of such numbers of \textit{certain} forms or characteristics . 

Number Field Sieve (NFS) algorithm is an example of general purpose algorithms where the running time depends solely on the size of integer to be factored. Continued fraction factorization and Quadratic sieve are some of other examples of general purpose factoring algorithms. 

The current fastest algorithm is the \textit{General Number Field Sieve} (GNFS), and it runs in approximately

$$
O\left(\exp \left(\frac{64}{9} n^{1 / 3}(\log n)^{2 / 3}\right)\right)
$$

time, where $n$ refers to bits used to represent the number. Shor's algorithm \cite{shor1994algorithms} runs in

$$
O\left((\log n)^{2} \cdot \log \log n\right)
$$

time \cite{GNFS-shor}. Even though Shor's algorithm presents an attack in polynomial time, it assumes a quantum computer free of noise and errors which is still a practical challenge.

In elliptic-curve based cryptography, Pollard $rho$ or baby-step giant-step method are some of the most efficient algorithms for solving ECDLP problem and needs $O(\sqrt{n})$ steps to solve, where $n$ is the order of the group. The size of the underlying field should be roughly twice the security parameter. For example, for 128-bit security one needs a 256-bit elliptic curve. In contrast, the size of key in RSA would be 3072 bits for the same level of security. Elliptic-curve based schemes offer shorter keys and hence, are widely used for protocols and cryptographic constructions. Choice of \textit{weaker} implementations of curve can affect the security and hence NIST Digital Signature Standards \cite{nistDSSfips} contains recommendations for curves that are chosen for efficiency and optimal security. 

\cleardoublepage

\chapter{Blind Signatures}
\label{ch:chapter_3}

Blind signatures are an extension of digital signatures, which allows a user to obtain a signature on a message without revealing the contents of the message to the signer. A good analogy to blind signatures is electronic voting. Though there are other security factors in implementations of e-voting in the real world \cite{appel2020democracy}, the one logically desired property in such scheme is anonymity i.e. the election authority should not be able to view any information about the votes. 

\begin{figure}
    \centering
    \includegraphics[scale=0.92]{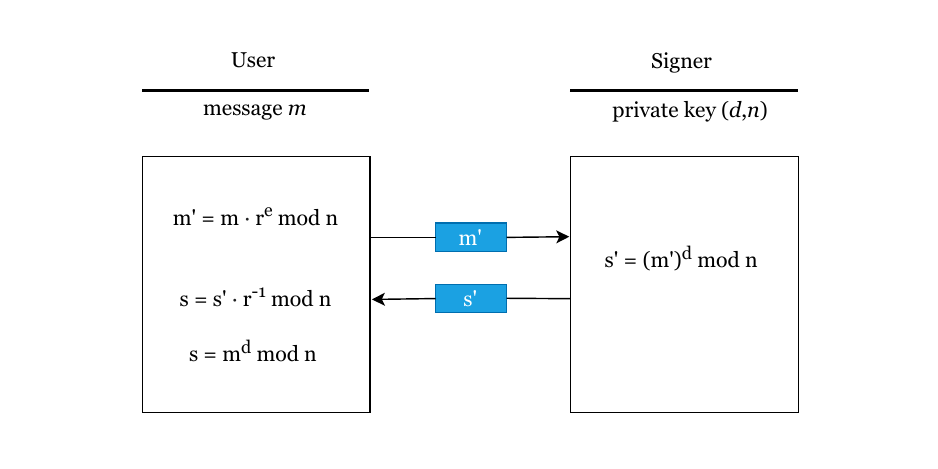}
    \caption{RSA blind signatures}
    \label{fig:rsa-blind-sing}
\end{figure}

To have a better idea, we describe the simple RSA blind signature scheme, a user who wants a message signed first blinds the message by multiplying it with a random blinding factor, \( r \), raised to the public exponent, \( e \), modulo the RSA modulus, \( n \). This blinded message, \( m' = m \cdot r^e \mod n \), is sent to the signer. The signer, unaware of the original message, signs the blinded message using their private key, \( d \), resulting in \( s' = (m')^d \mod n \). The user then receives the blinded signature, \( s' \), and removes the blinding factor by computing \( s = s' \cdot r^{-1} \mod n \), where \( r^{-1} \) is the modular inverse of \( r \). The result, \( s \), is a valid RSA signature on the original message \( m \), ensuring the user's message is signed without revealing the message to the signer, thus maintaining the user's privacy.

The correctness for the signature $s$ is as follows:

\[ s = (m \cdot r^e)^d \cdot r^{-1} \mod n \\ \]
\[ = m^d \cdot (r^e)^d \cdot r^{-1} \mod n \\ \]
\[ = m^d \cdot r^{ed} \cdot r^{-1} \mod n \\ \]
\[ = m^d \cdot r^{1} \cdot r^{-1} \mod n \quad \text{(since \( ed \equiv 1 \mod \phi(n) \))} \\ \]
\[ = m^d \cdot r \cdot r^{-1} \mod n =  m^d \mod n \]

Thus, \( s \) is indeed \( m^d \mod n \), which is the standard RSA signature of the original message \( m \).

\section{Blind signature scheme}

According to \cite{chaumblind}, a blind signature scheme $\Pi_{\mathrm{BS}}$ consists of three algorithms ($KeyGen$, $Sign$, $Verify$) and two parties, the user $U$ and the signer $S$:

\begin{description}
	\item [\textit{\textbf{KeyGen($k$)}}] It takes a security parameter $k$ as the input and generates a pair ($pk$,$sk$) as the output. $sk$ is the private key and $pk$ is the corresponding public key. We can denote this as KeyGen($k$) $\rightarrow$ ($sk$, $pk$)
 
	\item [\textbf{\textit{Sign($m,sk$)}}] This is an interactive and probabilistic polynomial time protocol between the user and signer where the user $U$ intends to get the signature $\pi$ on a message $m$ from the signer $S$. The user blinds the message $m$ to get $m'$ and sends it to the signer. With the private key $pk$, signer signs on $m'$, obtains $\pi'$ and sends it to the user. The user $U$ now unblinds $\pi'$ to get $\pi$, the blind signature on message $m$. This is denoted as Sign($m,sk$) $\rightarrow$ $\pi$
 
	\item [\textbf{\textit{Verify($m,\pi,pk$)}}] This deterministic  algorithm takes as input a message $m$, the signature $\pi$ to verify and the public key $pk$. It returns $true$ or $1$ if ($m$, $\pi$) is a valid message-signature pair. Otherwise, the signature is rejected and algorithm returns $false$ or 0. We can denote it as Verify($m$,$\pi$,$pk$)  $\rightarrow$ ($accept$/$reject$)
\end{description}

\section{Security}

The example of simple RSA blind signature which we saw earlier, is not secure. An adversary can use the blind signature process to trick a signer into decrypting a RSA ciphertext. An adversary, who has an encrypted message \( c = m^e \mod n \) and wants to decrypt it, can choose a random blinding factor $r$ and compute the blinded message \( c' = c \cdot r^e \mod n \). The adversary then presents \( c' \) to the signer for a blind signature. The signer, unaware that $c'$ is related to $c$, signs $c'$ using their private key $d$, resulting in \( s' = (c')^d \mod n \). The adversary receives $s'$, and by removing the blinding factor, computes \( s = s' \cdot r^{-1} \mod n \). Since \( s' = (c \cdot r^e)^d \mod n = m \cdot r \mod n \), removing $r$ reveals $m$, effectively decrypting the original message $c$.

Moreover, an adversary who wants to forge signature on a message $m$ can do so by requesting signatures for two other messages 

\[ {m_1} \mod n \] 
\[{m \cdot {m_1}^{-1}} \mod n \]

With their respective signatures $s_1$ and $s_2$, he can compute $s = s_1$ $\cdot $ $s_2 \mod n$ to obtain signature $s$ for the message $m$. 

We see the secure version of RSA blind signature scheme in the section \ref{sec:rsa-fdh} where the hash or \textit{digest} of the message $m$ is signed
. A secure blind signature scheme must satisfy the following three requirements \cite{han2005pairing}:

\begin{description}
    \item [Correctness:] The blind signature verification algorithm Verify($m$,$\pi$,$pk$) will always accept the signature $\pi$ if both, the signer $S$ and the user $U$ follow the signature generation correctly i.e. Verify($m$,$\pi$,$pk$)  $\rightarrow$ ($accept$)
    
    \item [Unforgeability:] The user $U$ should not be able to generate or forge signatures without the knowledge of signer's private key $sk$. More precisely, if the user is in possession of $l$ valid pairs of messages and signatures, then the signer $S$ ran the signature generation algorithm at least $l$ times. This property is also called 'one-more forgery' which was first mentioned in \cite{pointcheval}
    \item [Blindness:] Let $V$ be the view of the protocol that consists of all the parameters which are visible to the signer. While running one instance of blind signature scheme, for a valid message-signature pair of ($m$,$\pi$) and the corresponding view $V$, the signer $S$ cannot link this view $V$ to ($m$,$\pi$) at a later time. This is also called unlinkability \cite{chaumblind} and simply means that the signer's view $V$ of ($m$,$\pi$) and $V'$ of another pair, ($m'$,$\pi'$) are statistically independent.
\end{description}

\section{Evolution of blind signature schemes}
\label{sec:blind-evo}

After proposing the concept of blind signatures \cite{chaumblind} for the first time, Dr. D. Chaum also proposed the RSA-based blind signature scheme \cite{Chaum1984}. In 1994, Camenisch et al. proposed blind signature scheme based on discrete logarithm problem for the first time in \cite{Camenisch1995}. The paper presented two schemes in which one was derived from a variation of Digital Signing Algorithm (DSA) \cite{DSS1994}, the second one was based on the Nyeberg-Rueppel signature scheme \cite{Nyberg1994}. Pointcheval \cite{pointcheval} \cite{pointcheval1998} provided the first provably secure blind signature scheme.

In 1992, Solms et al. \cite{VONSOLMS1992} raised concerns regarding dangers of implementing blind signatures for digital cash/currency. The authors argued that the total anonymity can be misused to commit `perfect crime' where authority has no way of linking a transaction to criminals. They can use such payment systems to obtain ransoms or launder money. To address this issue, Stadler et al. \cite{stadler1995} proposed the concept of \textit{fair} blind signature scheme. With the help of a trusted third-party, the anonymity of a user can be revoked in case of a dispute between signer and user. The work proposed two schemes in which the revocation mechanism works in two separate directions: given the signer's view $V$ of a session, the third party could enable the signer to efficiently identify the message-signature pair, or links a given message-signature pair to the corresponding view $V$ of the protocol. 

This raises the point that in \textit{fully} blind signatures, the signers are oblivious to the message being signed. For example, in a simple e-cash system, the signer would like to control certain attributes like `date of issue' or `validity' rather than the user. \textit{Partially} blind signatures compensates for these features. Abe and Fujisaki \cite{abepartial} introduced the concept of partially blind signatures in which signer can include some information which is agreed upon earlier. In the example of e-cash, the signer can add details about the validity of a token in the signing message to prevent abuse. Abe and Okamoto \cite{abe_okamoto_partial} then gave a formal definition of such scheme and gave proof of completeness, partial blindness and unforgeability in the \textit{Random Oracle Model} (ROM).

\section{Blind signature schemes}

In this section, we go over the construction of blind RSA \textit{Full-Domain Hashing} scheme which is based on the computational difficulty of the RSA problem (and the related integer factorization problem) and \textit{Schnorr} blind signatures, whose securities relies on the discrete logarithm problem. 

\subsection{RSA-FDH Blind signature}
\label{sec:rsa-fdh}

This is the modified version of the RSA-FDH \cite{bellare1993random} signature scheme which follows the \textit{hash-and-sign} paradigm. It hashes the message onto the full domain of the RSA cryptosystem i.e. image(output) size of the hash function equals the size of the RSA modulus $N$.

\subsubsection*{Key generation}
In the RSA-FDH scheme, key generation follows the standard RSA key generation process. The signer selects two large prime numbers, \( p \) and \( q \), and computes their product \( n = pq \), which is used as the modulus. The public exponent \( e \) is chosen such that it is relatively prime to \( \phi(n) = (p-1)(q-1) \), and the private exponent \( d \) is computed as the modular inverse of $e$ mod \( \phi(n) \). The public key consists of \( (e, n) \) and the private key is \( (d, n) \).

\subsubsection*{Signing}
The signing process involves three steps: blinding, signing, and unblinding. First, the user chooses a random blinding factor $r$ such that \( \text{gcd}(r, n) = 1 \) and computes the blinded message \( m' = H(m) \cdot r^e \mod n \), where $H(m)$ is the digest of message $m$ from the hash function. \( H: \{0,1\}^* \rightarrow \mathbb{Z}_n \). The user then sends $m'$ to the signer. The signer, using their private key $d$, computes the blind signature \( s' = (m')^d \mod n \) and sends $s'$ back to the user. The user then unblinds the signature by computing \( s = s' \cdot r^{-1} \mod n \), resulting in a valid signature $s$ on the original hashed message $H(m)$.

\subsubsection*{Verification}
The verification process is straightforward and follows the standard RSA signature verification. The verifier, who knows the signer's public key $pk$ consisting\( (e, n) \), verifies the signature by checking that \( s^e \mod n = H(m) \). If this equation holds true, the signature is considered valid. This verification confirms that the signature was indeed produced by the signer's private key and that the message has not been altered, thus ensuring the integrity and authenticity of the signed message.

Completeness: If the user and signer follow the protocol honestly, then if 

\[ s' = (m')^d \mod n   \] 
\[ s = (s' \times r^{-1}) \mod n \rightarrow \text{ then $s$ is a valid signature for $m$ as }\]
\[  s^e \mod n = H(m) \]

Unforgeability: \cite{stadlerRSA98} showed that one-more forgery is possible if the hash function H is poorly implemented. However \cite{OneMoreRSA2001Bellare} introduced `\textit{one-more-RSA-inversion}' problem and gave proof that the scheme is unforgeable if hash function $H$ is considered a random oracle.

\subsection{DLP based blind Schnorr signatures}

One of the recommended digital signatures schemes by NIST \cite{nistDSSfips} is the Edwards-curve Digital Signature Algorithm (EdDSA). This scheme uses a variant of a Schnorr signature \cite{schnorr1990efficient} based on elliptic curves called \textit{twisted Edwards} curves. Before we look at the scheme of blind Schnorr signatures, we have a brief description \cite{demarmels2022adding} of the Schnorr \cite{schnorr1990efficient} signature scheme.

\subsubsection*{Schnorr signatures}

With group $\mathbb{G}$ of order $p$ and a generator $G$,  the private key is a randomly generated integer $x \stackrel{\$} \leftarrow \mathbb{Z}_p$. The public key is then $X \leftarrow x G$.

To sign a message $m$, signer chooses $r \stackrel{\$} \leftarrow   \mathbb{Z}_p$ and sends the value $R:=r G$ to the user. The user calculates $c:=H(R, m)$ with hash function $H:\{0,1\}^{*} \rightarrow \mathbb{Z}_p$ and sends the value of $c$ back. With $c$, signer finally calculates $s:=r+c x \bmod p$ and publishes the signature on message $m$ as $\sigma:=(R, s)$.

The user can evaluate the condition $s G=R+c X$ to check if $\sigma$ is a valid signature. The correctness holds because $s G=R+c X$ is $(r+c x) G=r G+c x G$.

\subsubsection*{Blind Schnorr signatures}

\subsubsection*{Key generation}

With group $\mathbb{G}$ of order $p$, generator $G$ a hash function $H:\{0,1\}^{*} \rightarrow \mathbb{Z}_p$, public parameters $(p, \mathbb{G}, G, H)$ are setup. Then key generation follows where the private key is a randomly generated integer $x \stackrel{\$} \leftarrow \mathbb{Z}_p$ and public key is then $X \leftarrow x G$.

\subsubsection{Signing: An interactive protocol}

The goal is to get the signature $\sigma:=( R^{\prime}, s^{\prime})$ for a message $m$. As before, signer chooses $r \stackrel{\$} \leftarrow  \mathbb{Z}_p$ and sends the value $R:=r G$ to the user. This time, the user generates two random blinding factors $\alpha, \beta \stackrel{\$} \leftarrow \mathbb{Z}_p$ and calculates $R^{\prime}:=R+\alpha G+\beta X$. 
User also computes $c^{\prime}:=H\left(R^{\prime}, m\right)$ and then uses $\beta$ to blind $c$ as $c:=c^{\prime}+\beta \bmod p$. The challenge $c$ is then blindly signed by the signer by doing $s:=r+c x \bmod p$. To get the unblinded signature, the user has to unblind $s$ as $s^{\prime}:=s+\alpha \bmod p$. And the signature is $\sigma:=( R^{\prime}, s^{\prime})$. 

\subsubsection*{Verification}
For the signature $\sigma:=( R^{\prime}, s^{\prime})$ where $X$ is the public key and $m$ is message. A user who wants to verify the signature can do so by checking if the following equation holds:
$$
\begin{aligned}
s^{\prime} G & =R^{\prime}+c^{\prime} X \\
& =R^{\prime}+H\left(R^{\prime}, m\right) X
\end{aligned}
$$

\subsubsection{Correctness}

For the verification, the following equation is checked

\begin{equation}
    \label{eq:schnorr-eq}
    s^{\prime} G =R^{\prime}+c^{\prime} X
\end{equation}

which is the same form as in original Schnorr signature, $s G=R+c X$

In eq. \ref{eq:schnorr-eq} we replace $s, R, c$, with the values used in the blind signature scheme.

$$
\begin{aligned}
s & =s^{\prime}-\alpha \\
R & =R^{\prime}-\alpha G-\beta X \\
c & =c^{\prime}+\beta
\end{aligned}
$$
and then the correctness holds for eq. \ref{eq:schnorr-eq} as:
$$
\begin{aligned}
s G & =R+c X \\
\left(s^{\prime}-\alpha\right) G & =R^{\prime}-\alpha G-\beta X+\left(c^{\prime}+\beta\right) X \\
s^{\prime} G-\alpha G & =R^{\prime}-\alpha G+c^{\prime} X \\
s^{\prime} G & =R^{\prime}+c^{\prime} X
\end{aligned}
$$

\section{Other types of blind signatures}

In section \ref{sec:blind-evo}, we mentioned three types of blind signature - total, fair and partially blind signatures. There are other variants that have studied and proposed by researchers over time. We go over them briefly: 

\begin{description}
    \item [Restrictive blind signatures] In a total blind signature scheme, the signer is totally unaware of the message content. Restrictive blind signature provides the desirable ability for the signer to impose certain restrictions or conditions on the blinded message before signing. For example, in e-cash system, the signer (bank) may require customers to include their identity in the blinded message.
    \item [Proxy blind signatures] Proxy blind signatures allow a designated proxy entity to obtain a blind signature on behalf of another user. The motivation is to \cite{liu2024survey} implement secure delegation of signature authority, that is, by introducing a proxy signer that can sign on behalf of the original signer, and the proxy cannot forge the signature of the original signer. 
    \item [Group blind signatures] Group blind signatures allow a group of users to collectively obtain a signature on a message in such a way that the signer cannot determine which user requested the signature. This ensures anonymity within the group while still providing a valid signature.
    \item [Threshold blind signatures] In threshold blind signature schemes, multiple signers collectively generate a blind signature for a message. A certain threshold of signers must collaborate to produce the signature, ensuring that no single signer can produce the signature alone. This enhances security and prevents any one entity from abusing the signing power.
    \item [Blind multi-signatures] Blind multi-signatures extend the concept of blind signatures to support multiple signers. Multiple parties can jointly sign a message without revealing individual signing keys or requiring each signer to see the entire message. This allows for distributed signing while maintaining privacy and unlinkability.
\end{description}

\section{Security models and Attacks}

\subsection{Random Oracle Model (ROM)}

The Random Oracle Model, introduced by Bellare and Rogaway in \cite{bellare1993random}, is a theoretical framework allowing to prove the security of hash-and-sign signature schemes. In this model, the hash function is seen as an oracle that outputs a random value for each new query. However, security proofs in the random oracle are not real proofs, since the random oracle is replaced by a well defined hash function in practice. Canetti et al. \cite{canetti2004random} showed that a security proof in the random oracle model does not necessarily imply that a scheme is secure in the real world.

In ROM, all the algorithms of the blind-signature scheme $K G, \mathcal{S}, \mathcal{U}, Vf$ gets a black-box access to an oracle $O:\{0,1\}^{n} \rightarrow\{0,1\}^{n}$, and can query the oracle on any input $x$. Initial constructions of blind signature schemes were in the random oracle model.

Pointcheval et al. \cite{pointcheval} were the first to give a secure blind signature schemes. They first proposed various definitions of unforgeability:

\begin{itemize}
    \item (The $(l, l+1)$-Forgery). For any integer $l$, an $(l, l+1)$-Forgery comes from an attacker that provides $l+1$ signatures after $l$ interactions with the signer $\mathcal{S}$.

    \item (The "One-More" Forgery). A "One-More" Forgery is an $(l, l+1)$ Forgery for some integer $l$, polynomially bounded in the security parameter $k$.

    \item (The Strong "One-More" Forgery ). A Strong "One-More" Forgery is an $(l, l+1)$-Forgery for some integer $l$, poly-logarithmically bounded in the security parameter $k$ i.e. $l \leq(\log k)^{\alpha}$ for some constant $\alpha$.
\end{itemize}

They also focus on two kinds of attacks -

\begin{itemize}
    \item The sequential attack : the attacker interacts sequentially with the signer. For example, in a e-cash system, this attack can be performed by a user who withdraws coins, one after the other.
    \item The parallel attack : the attacker interacts $l$ times in parallel with the signer. This attack is stronger. The attacker can initiate new interactions with the signer before previous ones have ended. In the same example of e-cash, this attack can be performed by a group of users who withdraw many coins at the same time.
\end{itemize}

Till 2000, known practical blind signature schemes whose security against adaptive and parallel attacks can be proven in the random oracle model either needed five data exchanges between the signer and the user or are limited to issue only logarithmically many signatures in terms of a security parameter. Abe \cite{abe2001secure} presented an efficient blind signature scheme that allows a polynomial number of signatures to be securely issued while only three data exchanges are needed.

The scheme given by Abe provides polynomial security, i.e., one-more unforgeable even if polynomially many signatures are issued in an adaptive and concurrent manner. Another advantage of the scheme is its potential support of protocols that need additional functionality. One can easily extend the scheme to be partially blind schemes. The security is proven in the random oracle model. The scheme remains practical as it requires only three to four times more computation than the original Schnorr signatures. 

\subsection{Standard Model}

In cryptography the standard model is the model of computation in which the adversary is only limited by the amount of time and computational power available. Cryptographic schemes are usually based on complexity assumptions, which state that some problem, for example integer factorization, cannot be solved in polynomial time. Schemes which can be proven secure using only complexity assumptions are said to be secure in the standard model. Security proofs are notoriously difficult to achieve in the standard model, so in many proofs, cryptographic primitives are replaced by idealized versions.

Juels et al. \cite{juels1997security} gave a blind signature scheme and this was the first scheme whose proof of security was proven only using complexity assumptions. All previous proofs were in random oracle model, based on number-theory assumptions only and were not fully polynomial. They showed how to achieve their protocol based on any one-way trapdoor permutations. Lindell also showed the impossibility of concurrently-secure blind signatures if simulation-based definitions of security are used \cite{lindell2003bounded}. However, Hazay et al. \cite{hazay2007concurrently} presented a concurrently-secure blind signature scheme which relies on standard cryptographic assumptions (e.g., trapdoor permutations and the decisional Diffie-Hellman assumption), and they prove security with respect to game-based definitions that are stronger, also bypassing the impossibility result of Lindell  \cite{lindell2003bounded}.

\subsection{Common Reference String model}

The common reference string (CRS) is a model where there is a public string that was generated in a trusted manner, and all parties have access to the string. Schemes proven secure in the CRS model are secure given that the setup was performed correctly. Datta \cite{datta2011survey} states the definition (\cite[Definition 3]{datta2011survey}):

\begin{definition} \textit{A blind signature scheme, in Common Reference String (CRS) model, consists of a tuple of effcient algorithms $B S=(C, K G,\langle\mathcal{S}, \mathcal{U}\rangle, V f)$ where}

\textit{\textbf{CRS Generation}. $C\left(1^{n}\right)$ generates a common reference string crs.}

\textit{\textbf{Key Generation} $K G(c r s)$ generates a key pair (sk,pk).}

\textit{\textbf{Signature Issuing} The joint execution of algorithm $\mathcal{S}(c r s, s k)$ and algorithm $\mathcal{U}(\mathrm{crs}, p k, m)$ for message $m \in\{0,1\}^{n}$ generates an output $\sigma$ of the user,}

$(\perp, \sigma) \leftarrow\langle\mathcal{S}(c r s, s k), \mathcal{U}(c r s, m, p k)\rangle$

\textit{\textbf{Verification} $Vf(c r s, p k, m, \sigma)$ outputs a bit.}

\end{definition}

We conclude that a scheme is complete if, for a valid signature $\sigma$, generated after the joint execution of $\mathcal{S}(c r s, \mathrm{sk})$ and $\mathcal{U}(c r s, p k, \mathrm{~m})$, the verification algorithm Vf $(c r s, p k, \mathrm{~m}, \sigma)$ outputs the bit 1.

The main advantage of the CRS model over the random oracle model is that security is standard, and doesn't rely on a heuristic belief system that the real protocol that uses a standard hash function is secure. In \cite{fischlin2006round} Fischlin presented a scheme in the CRS model which has a two-move signature protocol. The scheme is a generic construction from basic primitives, namely schemes for commitment, encryption and signatures as well as generic non-interactive zero knowledge (NIZK) proofs for NP-languages. The signature request protocol consists of the user sending a commitment to the message to the signer, who responds with a signature on the commitment. The user then uses this signature on the commitment to construct the blind signature, by first encrypting the commitment and the signature, and then adding a NIZK proof that the encrypted signature is a valid signature on the encrypted commitment. Such two-move signature scheme is round-optimal (single message by user and single response from signer). It doesn't deal with the issue of security in concurrent setting and with the desire to avoid the use of the random oracle, much recent work has focused on developing round-optimal blind signatures in the CRS model.

\subsection{ROS Attack}
\label{sec:ros}

The $ROS$-problem was originally studied by Schnorr \cite{schnorr2001security} in the context of blind signature schemes. They presented a novel parallel one-more signature forgery against blind Okamoto-Schnorr and blind Schnorr signatures in which an attacker interacts some $l$ times with a legitimate signer and produces from these interactions $l+1$ signatures. The attack uses a solution of the following ROS-problem which Schnorr \cite{schnorr2001security} states as follows:

\textbf{ROS-problem:} Find an overdetermined, solvable system of linear equations modulo $q$ with random inhomogenities. Specifically, given an oracle random function $F: \mathbf{Z}_q^l \rightarrow {\mathbf{Z}}_q$, find coefficients $a_{k, \ell} \in \mathbf{Z}_q$ and a solvable system of $l+1$ distinct equations in the unknowns $c_1, \ldots, c_l$ over $\mathbf{Z}_q$ :
$$
a_{k, 1} c_1+\ldots+a_{k, l} c_l=F\left(a_{k, 1}, \ldots, a_{k, l}\right) \text { for } k=1, \ldots, t .
$$

They presented an attack against Schnorr signatures which does not depend on the generator $g$ or the public key $h$. 

The signer sends commitments $g_1=$ $g^{r_1}, \ldots, g_l=g^{r_l}$. The attacker $\mathcal{A}$ selects $a_{k, 1}, \ldots, a_{k, l} \in \mathbf{Z}_q$ and messages $m_1, \ldots, m_t$, and computes $f_k=g_1^{a_{k, 1}} \cdot \ldots \cdot g_l^{a_{k, l}}$ and $H\left(f_k, m_k\right)$ for $k=1, \ldots, t$. The coefficients $a_{k, \ell}$ selected by the attacker are arbitrary values. 

Then $\mathcal{A}$ solves $l+1$ of the $t$ equations in the unknowns $c_1, . ., c_l$ over $\mathbf{Z}_q$ :

\begin{equation}\label{ros-eq}
    H\left(f_k, m_k\right)=\sum_{\ell=1}^l a_{k, \ell} c_{\ell} \text { for } k=1, \ldots, t .
\end{equation}

$\mathcal{A}$ sends the solutions $c_1, \ldots, c_l$ as challenges to the signer. The signer sends back $z_{\ell}:=r_{\ell}+c_{\ell} x \in \mathbf{Z}_q$ for $\ell=1, . ., l$. For each solved equation \ref{ros-eq}, the attacker gets a valid signature $\left(m_k, c_k^{\prime}, z_k^{\prime}\right)$ by setting
$$
c_k^{\prime}:=\sum_{\ell=1}^l a_{k, \ell} c_{\ell}=H\left(f_k, m_k\right) \text { and } z_k^{\prime}:=\sum_{\ell=1}^l a_{k, \ell} z_{\ell} .
$$

Correctness holds as the equations \ref{ros-eq} imply that
$$
g^{z_k^{\prime}} h^{-c_k^{\prime}}=g_1^{a_{k, 1}} \cdot \ldots \cdot g_l^{a_{k, l}}=f_k \text { and } H\left(g^{z_k^{\prime}} h^{-c_k^{\prime}}, m_k\right)=c_k^{\prime} \text {. }
$$

Using a solver for the ROS problem, Wagner \cite{wagner2002generalized} showed that the unforgeability of the Schnorr and Okamoto-Schnorr blind signature schemes can be attacked in subexponential time whenever more than polylog $(\lambda)$ signatures are issued concurrently with $\lambda$ the security parameter. It was only recently that Benhamouda et al. \cite{benhamouda2021security} proposed an elegant polynomial time attack against $\mathrm{ROS}_{\ell}$ for $\ell=$ poly $(\lambda)$ concurrent sessions. A number of schemes are affected, it includes threshold signature scheme \cite{gennaro2007secure}, multi-signature scheme like \cite{syta2016keeping}, partially blind signature \cite{abe_okamoto_partial} and conditionally blind signature \cite{grontas2019towards}. Benhamouda et al. have shared such schemes in their work \cite{benhamouda2021security}. Their attack is very practical, for instance when $\ell=128$, it only takes time roughly $2^{32}$ hash computations to break unforgeability.

In \cite{katsumata2024breaking}, authors presented a problem called Parallel Random inhomogeneities in an Overdetermined Solvable system of linear equations (pROS) problem and show that an attack against pROS implies an attack to more blind signatures. The attack breaks 4-concurrent unforgeability of isogeny-based (partially) blind signature scheme, $CSI-Otter$ \cite{katsumata2023csi} in time roughly $2^{34}$ hash computations. Furthermore, it can achieve the same result for $Blaze+$ \cite{alkeilani2020lattice} and $BlindOR$ \cite{alkeilani2021blindor} with approximately $2^{43}$ hash computations.
\cleardoublepage

\chapter{Lattice-based Signatures}
\label{ch:chapter_4}

\section{Basics}

A lattice is a set of points in $n$-dimensional space with a regular structure. Consider the following matrix $\mathcal{A}$ with dimension $m \times n$. It is denoted as:

$$
\mathcal{A}=\left[\begin{array}{ccccc}
a_{11} & a_{12} & a_{13} & \ldots & a_{1 n} \\
a_{21} & a_{22} & a_{23} & \ldots & a_{2 n} \\
\vdots & \vdots & \vdots & \ddots & \vdots \\
a_{m 1} & a_{m 2} & a_{m 3} & \ldots & a_{m n}
\end{array}\right]
$$

Given $n$ linear independent vectors $\vec{b}_{1}, \ldots, \vec{b}_{n} \in \mathbb{R}^{m}$ with $m \geqslant n$, the generated lattice is formally defined to be the set of vectors:

$$
\mathcal{L}\left(\overrightarrow{\boldsymbol{b}}_{1}, \ldots, \overrightarrow{\boldsymbol{b}}_{\boldsymbol{n}}\right)=\left\{\sum_{i=1}^{n} a^{i} b^{i} \mid a^{i} \in \mathbb{Z}\right\}
$$

\begin{figure}[h]
    \centering
    \includegraphics[scale=0.30]{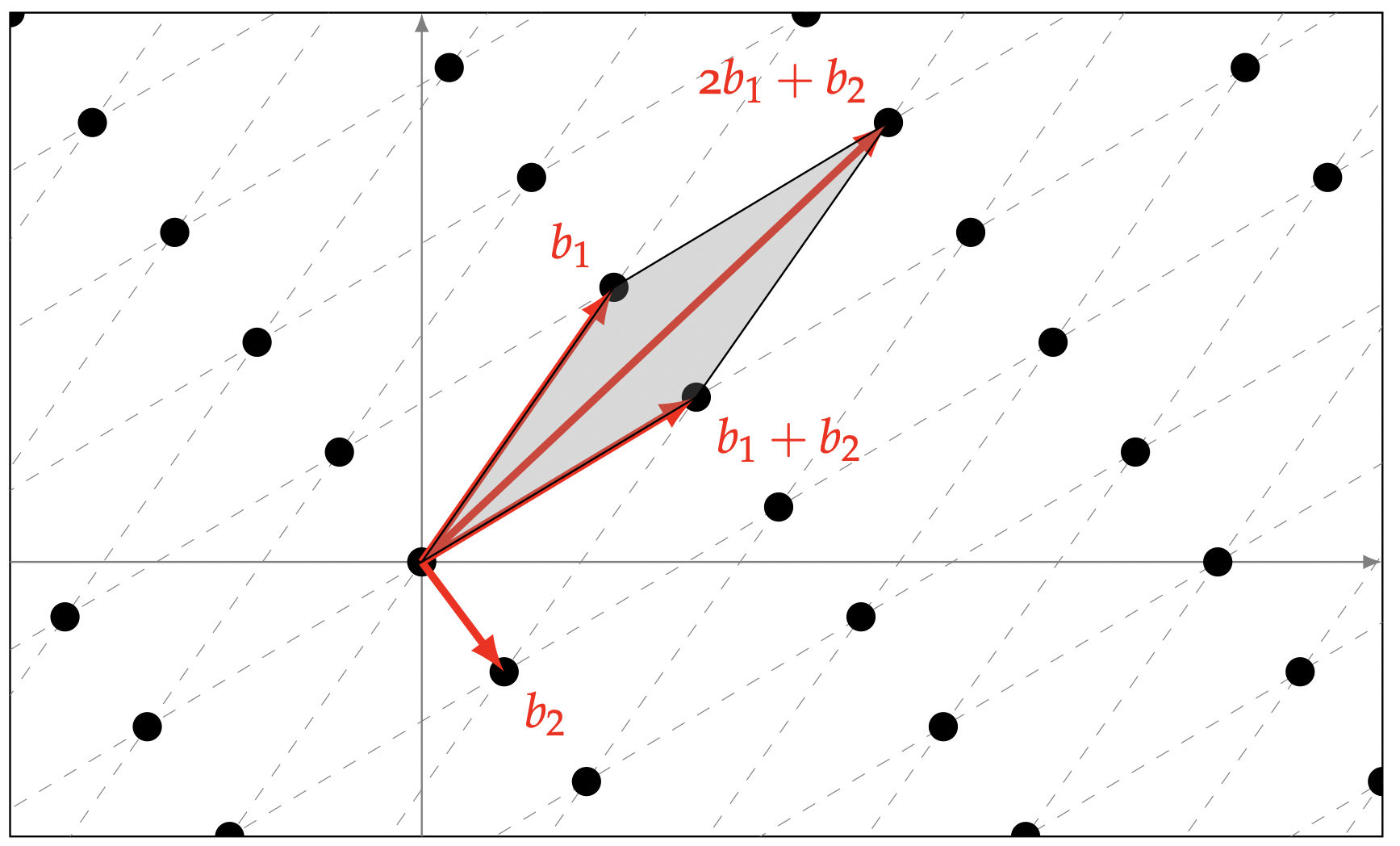}
    \caption{Lattice from the linear combinations of points $\vec{b}_{1}$ $and$ $\vec{b}_{2}$} \cite{varmasurvey}
    \label{fig:lattice-1}
\end{figure}

where vectors $\vec{b}_{1} \ldots, \vec{b}_{n}$ are called the basis of the generated lattice. The integers $n$ and $m$ are called rank and dimension respectively. In this thesis, we will only consider lattices where the rank and dimension are equal $(m=n)$.

All integer linear combinations of the vectors $\vec{b}_{1}, \ldots, \vec{b}_{n}$ generate a lattice. Note that the same lattice can be generated in several different ways. That is, the same lattice can be generated by different bases. Figure \ref{fig:lattice-1} gives an example of a 2-dimensional lattice with basis $\left\{\vec{b}_{1}, \vec{b}_{2}\right\}$. Combinations of linear vectors $\vec{b}_{1}+\vec{b}_{2}$ and $2 \vec{b}_{1}+\vec{b}_{2}$ generate the same lattice points.

Now we will define fundamental domain of a lattice and later see how can we measure how orthogonality of a lattice basis. This will be later important to understand good and bad basis. We refer to the work \cite{silverman2008introduction} to see the definition of fundamental domain.

\begin{definition}
Fundamental domain of $\mathcal{L}$: Suppose $\mathcal{L} \subset \mathbb{R}^n$ is an n-dimensional lattice with basis $\mathcal{B}=\left\{\vec{v}_1, \ldots, \vec{v}_n\right\}$. Then the fundamental domain $\mathcal{F}$ of $\mathcal{L}$ corresponding to $\mathcal{B}$ is:
$$
\mathcal{F}\left(\vec{v}_1, \ldots, \vec{v}_n\right)=\left\{t_1 \vec{v}_1+\ldots+t_n \vec{v}_n \mid 0 \leq t_i<1\right\}
$$    
\end{definition}

An important result regarding fundamental domains is the Hadamard's inequality \cite{silverman2008introduction} which states for fundamental domain $\mathcal{F}$ of lattice $\mathcal{L}$.
$$
\operatorname{det}(\mathcal{L})=\operatorname{Vol}(\mathcal{F}) \leq\left\|\vec{v}_1\right\| \ldots\left\|\vec{v}_n\right\|
$$

If $\mathcal{L}$ 's basis is orthogonal then the above is an equality. Thus the above can be seen as a measure of how orthogonal a basis of a lattice $\mathcal{L}$ is.

\begin{figure}[H]
    \centering
    \includegraphics[width=9cm, height=6cm]{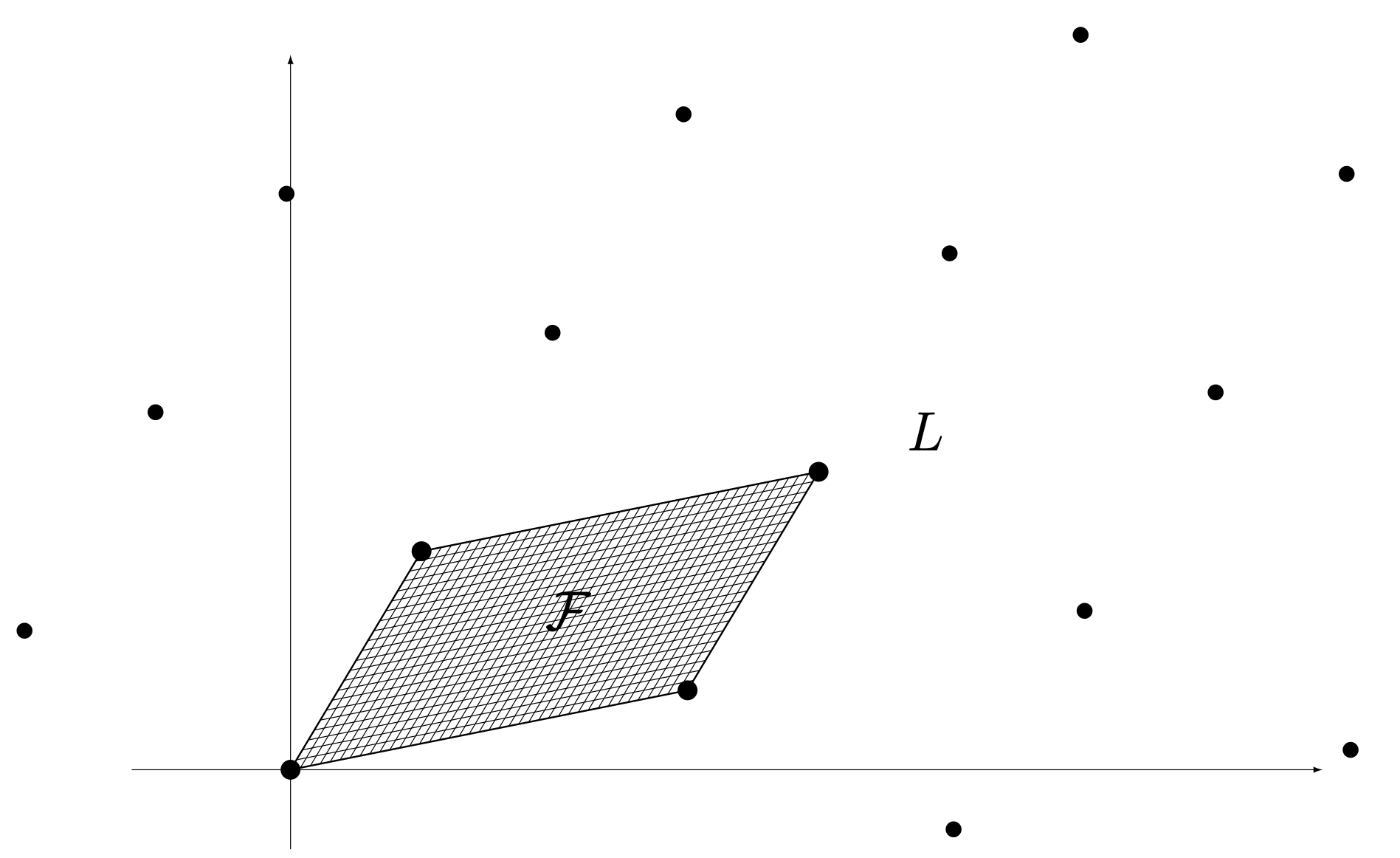}
    \caption{A 2-Dimensional lattice with fundamental domain $\mathcal{F}$ \cite{silverman2008introduction}}
    \label{fig:lattice-domain}
\end{figure}

\section{Hard problems in Lattices}

Although lattice structures are simple to define, there are some mathematical hard problems that form the basis of lattice-based cryptography.
Finding the shortest non-zero vector in a lattice is the most prominent one. The problem is to find a vector achieving the first lattice minimum \(
\lambda_1(\mathcal{L}):=\min _{\mathbf{v} \in \mathcal{L} \backslash\{\mathbf{0}\}}\|\mathbf{v}\|
\). This is the minimum distance of a lattice $\mathcal{L}$ which is the length of a shortest nonzero vector in the lattice. We refer to the work by Peikert \cite[Definitions 2.2.1, 2.2.2, 2.2.4]{peikert2016decade} which gives us following definitions.

\begin{definition}
Shortest Vector Problem SVP: Given an arbitrary basis $\mathcal{B}$ for lattice $\mathcal{L}$, find a shortest non-zero lattice vector (i.e. $v \in \mathcal{L}$ for which $\|v\|=\lambda_1(\mathcal{L})$ ).    
\end{definition}

\begin{definition}
Approximate Shortest Vector Problem $\mathrm{SVP}_\gamma$:  Given a basis $\mathcal{B}$ for a lattice $\mathcal{L}$, find a non-zero vector $v \in \mathcal{L}$ such that $\|v\| \leq \gamma(n) \cdot \lambda_1(\mathcal{L})$ (where $\gamma(n) \geq 1$ is an approximation factor taken to be a function of the lattice dimension $n)$.
\end{definition}

As the name suggests, $\mathrm{SVP}_\gamma$ asks us to find a lattice vector that's ``approximately'' the shortest (lattice) vector. A variant of $\mathrm{SVP}_\gamma$ is $\mathrm{SIVP}_\gamma$.

\begin{definition}
Approximate Shortest Independent Vector Problem SIVP$_\gamma$:  Given a basis $\mathcal{B}$ of $n$-dimensional lattice $\mathcal{L}$, output a set $\mathcal{S} \subset \mathcal{L}$ of $n$ linearly independent lattice vectors such that $\left\|s_i\right\| \leq \gamma(n) \cdot \lambda_n(\mathcal{L})$ for all $s_i \in \mathcal{S}$.    
\end{definition}

In high dimensional lattices, these problems becomes harder to solve. In 1996 Ajtai proved that the SVP is NP-hard for randomized reductions \cite{ajtai1996generating}. The Closest Vector Problem (CVP) is a computational problem on lattices closely related to SVP. Given a lattice $\mathcal{L}$ and a target point $t$, CVP asks to find the lattice point closest to the target. We define $dist$ which takes as input a n-dimensional lattice $\mathcal{L}$ and a target vector $t$ $\in$ $\mathbb{R}^n$ and returns dist(t,$\mathcal{L}$) := $\min _{\mathbf{v} \in \mathcal{L}}\|v\|$. Then CVP can be described as follows \cite[Definition 2.2]{wang2023lattice}: 

\begin{definition}
Closest Vector Problem (CVP): Given a lattice basis $\mathcal{L}$ and a target point $t$ $\in$ $\mathbb{R}^n$, find $v$ $\in \mathcal{L}$ such that
$$
\|v-t\|=\operatorname{dist}(t, \mathcal{L})
$$    
\end{definition}

In lattice-based cryptography, the exact SVP and CVP are rarely used as the security foundation directly. Most of lattice-based schemes actually correspond to some variants of SVP and CVP.

\section{Lattice-based cryptography}

The SVP and CVP problems are not convenient for straightforward constructions of lattice-based schemes. Some design-friendly hard problems, including SIS, LWE and NTRU, were later introduced for lattice-based cryptography. These problems and their variants have been the foundation of modern lattice-based cryptography. Next we give some preliminary descriptions to these problems.

\textbf{\textit{SIS (Short Integer Solution) problem}}:  The SIS problem was introduced and studied in Ajtai's breakthrough work \cite{ajtai1996generating}. Ajtai showed that the average-case SIS problem is at least as hard as the worst-case approximate SVP problem. SIS is the first hard problem for lattice-based cryptography of the worst-case/average-case reduction, which provides a strong security guarantee for lattice-based schemes. Since its introduction, SIS has been used as the foundations of many lattice-based primitives, e.g. hash functions and digital signatures.

\textit{SIS Problem}: For integers $n, m, q>0$ and a real number $B>0$, the $\operatorname{SIS}_{n, m, q, B}$ problem is as follows \cite[Definition 2.7]{wang2023lattice}: 

Given uniformly random $A \in \mathbb{Z}_{q}^{n \times m}$, find some non-zero $x \in \mathbb{Z}^{m}$ such that $\|x\| \leq B$ and $A x=0 \bmod q$.

\textbf{\textit{LWE (Learning With Errors) problem}}: Another cornerstone of lattice-based cryptography is the introduction of the LWE problem by Oded Regev \cite{regev2009lattices}. The average-case LWE can also be shown to be at least as hard as some worst-case lattice problems. Different from the SIS problem, LWE is able to be used to construct lattice-based public key encryption and cryptosystems. The LWE problem has a decision and search variant which \cite[Definition 2.8]{wang2023lattice} states as follows :

\textit{LWE Problem}: For integers $n, m, q>0$ and a distribution $\chi$ over $\mathbb{Z}$, let $A_{\mathbf{s}, \chi}$ for given $\mathbf{s} \in \mathbb{Z}^{n}$ be the distribution of $(a, b=\langle a, s \rangle+e \bmod q)$ where $a \leftarrow U\left(\mathbb{Z}_{q}^{n}\right)$ and $e \leftarrow \chi$.

- The decision $\mathrm{LWE}_{n, m, q, \chi}$ problem is as follows: Given $m$ samples drawn from $A_{s, \chi}$ where $s \leftarrow U\left(\mathbb{Z}_{q}^{n}\right)$ and $m$ samples from $U\left(\mathbb{Z}_{q}^{n} \times \mathbb{Z}_{q}\right)$, distinguish them.

- The search $\mathrm{LWE}_{n, m, q, \chi}$ problem is as follows: Given $m$ samples drawn from $A_{\mathbf{s}, \chi}$ where $s \leftarrow U\left(\mathbb{Z}_{q}^{n}\right)$, find $\mathbf{s}$.

\subsubsection{Lattice reduction}

\begin{figure}[h]
    \centering
    \includegraphics[width=10cm, height=6.5cm]{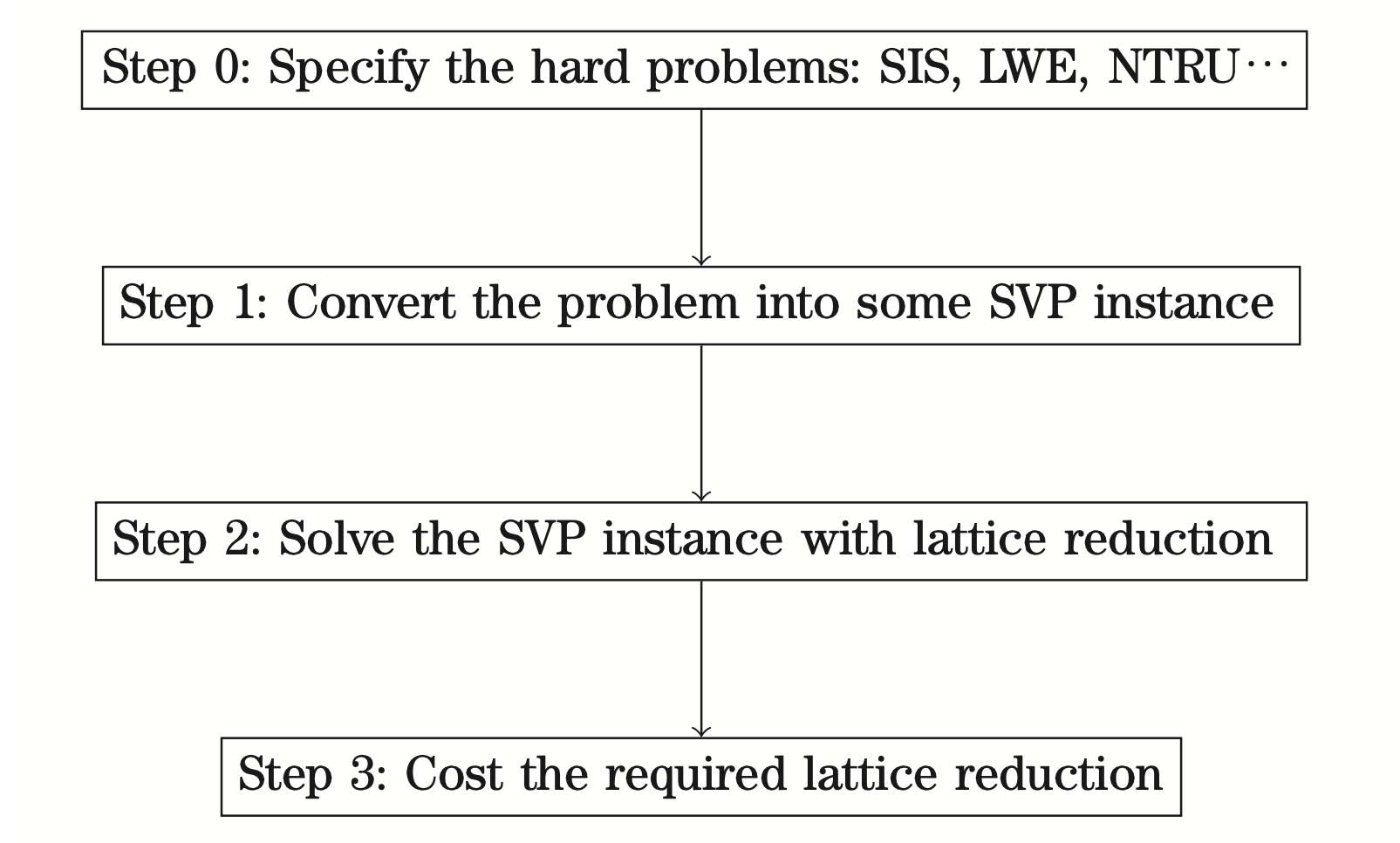}
    \caption{Cryptanalysis of lattice-based scheme \cite{wang2023lattice}}
    \label{fig:lattice-reduction}
\end{figure}

Cryptanalysis allows us to perform analysis of the security of cryptosystems. It avoids weak design and helps fix concrete security estimates and  parameter selections.  The security of a scheme relies on solving the underlying hard problems of the cryptosystem. Figure \ref{fig:lattice-reduction} shows steps of cryptanalysis of a lattice-based cryptosystem. 

There are two kinds of SVP algorithms \cite{yasuda2021survey}, depending on whether the give and exact solution or an approximate one:

- Exact SVP algorithms: They find the non-zero shortest lattice vector by searching through all the short vectors. Examples include enumeration and sieving.

- Approximate SVP algorithms: They output short lattice vectors, not always the shortest ones. These are faster than exact SVP algorithms. Examples include LLL and BKZ.

Now we refer to \cite[Definition 2.1]{deng2016introduction} for the definition of lattice reduction and also see an example of good and bad basis.

\begin{figure}[H]
    \centering
    \includegraphics[width=10cm, height=6.5cm]{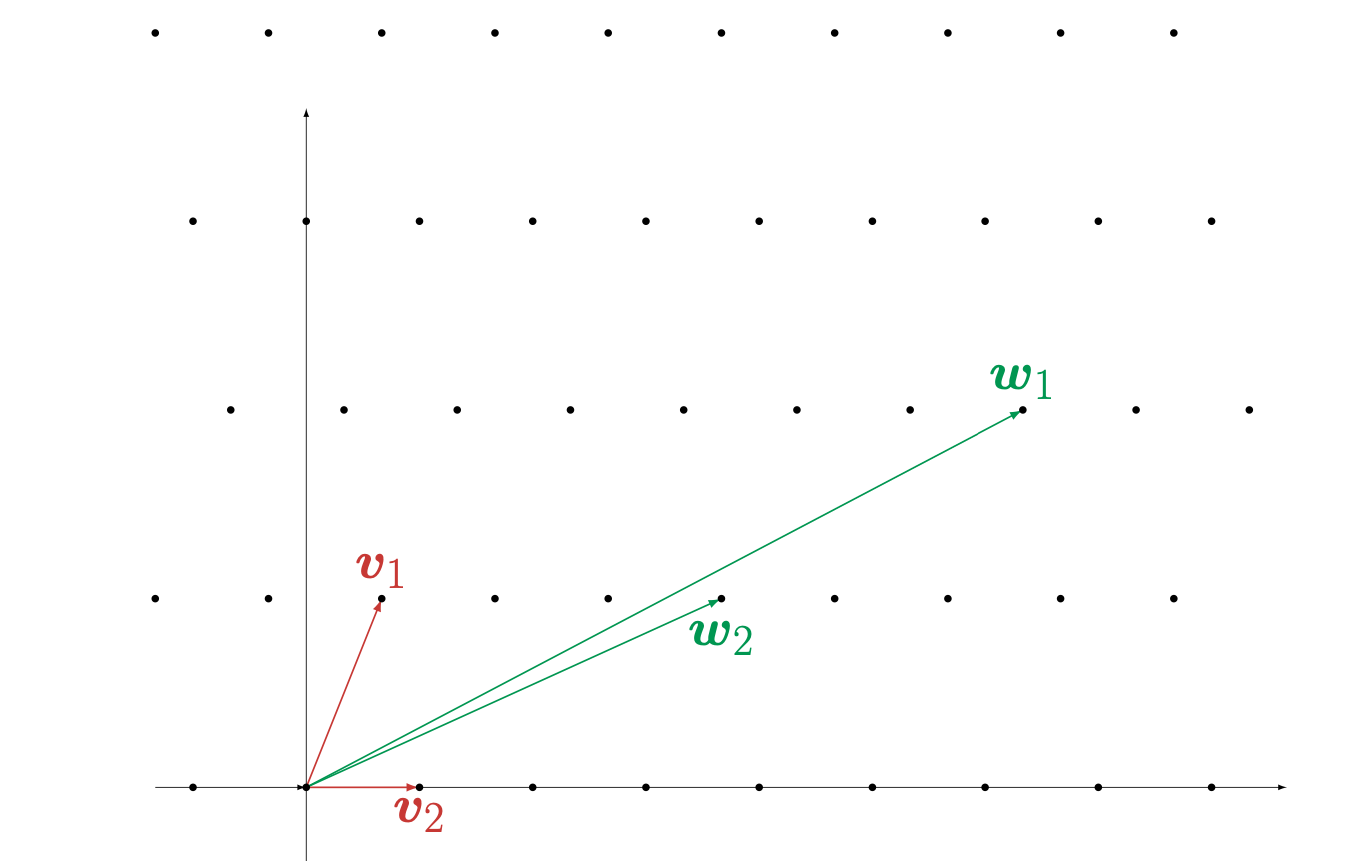}
    \caption{A `good' basis $\left\{\boldsymbol{v}_1, \boldsymbol{v}_2\right\}$ and a `bad' basis $\left\{\boldsymbol{w}_1, \boldsymbol{w}_2\right\}$}\cite{lattice_red_ppt}
    \label{fig:good-bad-basis}
\end{figure}

\begin{definition} 
    Lattice reduction is a process of reducing the basis $\mathbf{B}$ of a lattice $\mathcal{L}$ to a shorter basis $\mathbf{B}^{\prime}$ while keeping $\mathcal{L}$ the same. Figure \ref{fig:good-bad-basis} shows a lattice with two different basis. A basis $\left(\mathbf{b}_{\mathbf{1}}, \mathbf{b}_{\mathbf{2}}\right)$ is said to be reduced if it satisfies following condition:
$$
\begin{array}{r}
\left\|\mathbf{b}_{\mathbf{1}}\right\| \leq\left\|\mathbf{b}_{\mathbf{2}}\right\| \\
\end{array}
$$
$$
\begin{array}{r}
\displaystyle u=\frac{\mathbf{b}_{\mathbf{1}} \cdot \mathbf{b}_{\mathbf{2}}}{\left\|\mathbf{b}_{\mathbf{1}}\right\|^2} \leq \frac{1}{2} 

\end{array}
$$

$u$ is called the orthogonal projection coefficient
\end{definition}

\begin{figure}[H]
    \centering
    \includegraphics[height=8cm]{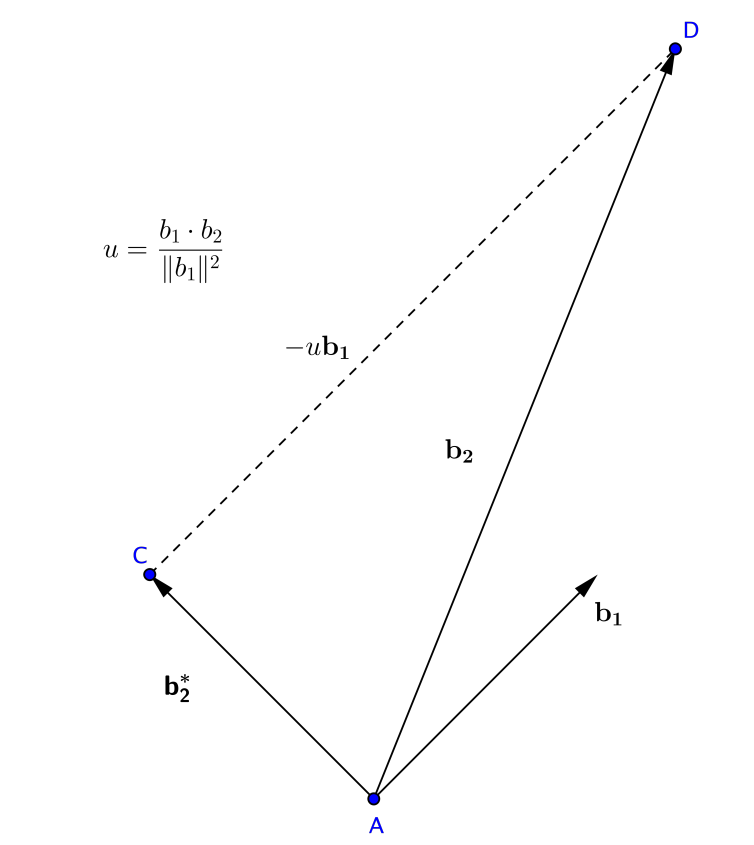}
    \caption{The set $\left\{\mathbf{b}_{\mathbf{1}}, \mathbf{b}_{\mathbf{2}}^*\right\}$ is the orthogonal basis for the lattice generated by basis $\left\{\mathbf{b}_{\mathbf{1}}, \mathbf{b}_{\mathbf{2}}\right\}$}\cite{deng2016introduction}
    \label{fig:Orthogonal-basis}
\end{figure}

\subsubsection{Gram-Schmidt Orthogonalization}

The idea of basis reduction in two dimensional lattice is to find the orthogonal basis based on the given basis \cite{deng2016introduction}. Gauss lattice reduction \cite{gauss_lattice} is one method of of finding optimal basis but only in two-dimensional lattice. We need to generalize the algorithm to n-dimensions. Gram-Schmidt Orthogonalization is also one process that generates orthogonal basis vectors and we refer to \cite[Theorem 3.1]{deng2016introduction} for the theorem.

\begin{theorem}
    Gram-Schmidt Orthogonalization: Given a basis $\left\{\mathbf{b}_{\mathbf{1}}, \mathbf{b}_{\mathbf{2}}, \cdots, \mathbf{b}_{\mathbf{m}}\right\}$ of a subspace $H_m$ of $\mathbb{R}^n$, we define
$$
\begin{aligned}
\mathbf{b}_{\mathbf{1}}{ }^* & =\mathbf{b}_{\mathbf{1}}, & \\
\mathbf{b}_{\mathbf{2}}{ }^* & =\mathbf{b}_{\mathbf{2}}-u_{1,2} \mathbf{b}_{\mathbf{1}}, & \text { where } u_{1,2}=\frac{\mathbf{b}_{\mathbf{2}} \cdot \mathbf{b}_{\mathbf{1}}{ }^*}{\mathbf{b}_{\mathbf{1}}{ }^* \cdot \mathbf{b}_{\mathbf{1}}{ }^*} \\
\vdots & & \\
\mathbf{b}_{\mathbf{m}}{ }^* & =\mathbf{b}_{\mathbf{m}}-\sum_{i<m} u_{i, m} \mathbf{b}_{\mathbf{i}}, & \text { where } u_{i, m}=\frac{\mathbf{b}_{\mathbf{m}} \cdot \mathbf{b}_{\mathbf{i}}{ }^*}{\mathbf{b}_{\mathbf{i}}{ }^* \cdot \mathbf{b}_{\mathbf{i}}{ }^*}
\end{aligned}
$$

Then, $\left\{\mathbf{b}_{\mathbf{1}}{ }^*, \mathbf{b}_{\mathbf{2}}{ }^*, \cdots, \mathbf{b}_{\mathbf{m}}{ }^*\right\}$ is an orthogonal basis of $H_m$.

\end{theorem}

\subsubsection{LLL Reduction}
Lenstra, Lenstra, and Lovász (and hence the name LLL algorithm) proposed an approximation algorithm \cite{lenstra1982factoring} using the Gram-Schmidt orthogonalization for lattice basis reduction in higher dimensions. Given a basis $\left\{\mathbf{b}_1, \mathbf{b}_2, \cdots, \mathbf{b}_{\mathbf{n}}\right\}$ where $n$ is the dimension, the LLL algorithm can be depicted as follows. We refer to \cite{deng2016introduction} for the algorithm steps.
 
\begin{algorithm}[H]
\caption{LLL Algorithm} 
\label{alg:LLL}
\begin{algorithmic}
\State Input :$\left\{\mathbf{b}_1, \mathbf{b}_2, \cdots, \mathbf{b}_{\mathbf{n}}\right\}$
\State Repeat two steps until find the LLL reduced basis
\State Step 1: Gram-Schmidt orthogonalization
\For{$i=1$ to $n$} 
        \For{$k=i-1$ to 1}  
              \State $m \leftarrow$ nearest integer of $u_{k, i}$ 
              \State $\mathbf{b}_{\mathbf{i}} \leftarrow \mathbf{b}_{\mathbf{i}}-m \mathbf{b}_{\mathbf{k}}$
        \EndFor
\EndFor
\State Step 2: Check Condition 2, and swap
\For{$i=1$ to $n-1$}
        \If {$\left\|\mathbf{b}_{\mathbf{i}+1}{ }^{*}+u_{i, i+1} \mathbf{b}_{\mathbf{i}}{ }^{*}\right\|^{2}<\frac{3}{4}\left\|\mathbf{b}_{\mathbf{i}}{ }^{*}\right\|^{2}$}
        \State swap $b_{i+1}$ and $b_{i}$
        \State go to step 1
        \EndIf
\EndFor

\end{algorithmic}
\end{algorithm}

They gave the definition of LLL reduced basis and we refer to \cite[Definition 4.1]{deng2016introduction} for the same:

\begin{definition}
    LLL reduced basis: Let $\left\{\mathbf{b}_1, \mathbf{b}_2, \cdots, \mathbf{b}_{\mathbf{n}}\right\}$ be a basis for a n-dimensional Lattice $\mathcal{L}$, and $\left\{\mathbf{b}_{\mathbf{1}}{ }^*, \mathbf{b}_{\mathbf{2}}{ }^*, \cdots, \mathbf{b}_{\mathbf{n}}{ }^*\right\}$ be the orthogonal basis generated due to Gram-Schmidt Orthogonalization, and we have $u_{i, k}=\frac{\mathbf{b}_{\mathbf{k}} \cdot \mathbf{b}_i{ }^*}{\mathbf{b}_{\mathbf{i}} \cdot \mathbf{b}_{\mathbf{i}}{ }^*}$. We say $\left\{\mathbf{b}_{\mathbf{1}}, \mathbf{b}_{\mathbf{2}}, \cdots, \mathbf{b}_{\mathbf{n}}\right\}$ is a LLL reduced basis if it satisfies two conditions:

(1) $\forall i \neq k, u_{i, k} \leq \frac{1}{2}$

(2) For each $i,\left\|\mathbf{b}_{\mathbf{i}+\mathbf{1}}{ }^*+u_{i, i+1} \mathbf{b}_{\mathbf{i}}{ }^*\right\|^2 \geq \frac{3}{4}\left\|\mathbf{b}_{\mathbf{i}}{ }^*\right\|^2$.
\end{definition}

The authors, Lenstra, Lenstra, and Lovász showed the result that LLL algorithm is a polynomial-time running algorithm and a LLL reduced basis is a good basis. \cite{gauss_lattice}. 

\section{Lattice trapdoors}

Let's consider the RSA function $f(x)=x^e \bmod N$ where $N=pq$ with the primes $p, q$ and $\operatorname{gcd}(e, \varphi(N))=1$. Then $d=e^{-1} \bmod \varphi(N)$ is the $trapdoor$ because $\left(x^e\right)^d=x \bmod N$. So if one can factorize $N$ into $p$ and $q$, then they can also calculate the trapdoor $d$. Thus, the function $f(x)$ is a trapdoor function because, given $f(x)$ and the secret trapdoor $d$, we can compute $x$.

Piekert \cite{peikert2016decade} states ``\textit{A trapdoor function is a function that is easy to evaluate and hard to invert on its own, but which can be generated together with some extra $trapdoor$ information that makes inversion easy}''. 

Gentry et al. \cite{GPV08} showed that we can use lattice problems to construct certain types of trapdoor functions. We now share two notions of lattice trapdoor as discussed by Vaikuntanathan in \cite[Definition 1.2, 1.4]{vekunt}:

\begin{definition}
\label{def:type1}
(`\textit{Type 1}' Trapdoor). For matrices $A \in \mathbb{Z}_{q}^{n} \times m$ and $T \in \mathbb{Z}_{q}^{m \times m}, T$ is a trapdoor for $A$ if:

1. $A T \equiv 0^{n \times m} \bmod q$

2. $T$ is full rank over $\mathbb{Z}$ i.e. rank is equal to the matrix dimension size

3. Each column $t_{i}$ of $T=\left[t_{1} t_{2} \cdots t_{m}\right]$ is short.
\end{definition}

The first type of trapdoor is a $short$ basis which is a lattice basis consisting of short-orthogonal vectors. This short basis can be used for efficient signing operations, where a hash of the message is mapped to a close lattice vector which is preimage-resistant without the trapdoor. The work in \cite{GPV08} also shared an approach to signing that provably leaks no information about the secret basis.

\begin{definition}
(`\textit{Type 2}' Trapdoor). For matrices $A \in \mathbb{Z}_{q}^{n} \times m$ and $R \in \mathbb{Z}_{q}^{m \times n \log q}, R$ is a trapdoor for $A$ if:

1. $A R=G$

2. Each column $r_{i}$ of $R=\left[r_{1} r_{2} \cdots r_{m}\right]$ is `short'.
\end{definition}

Here $G$ is a gadget matrix. It is a special matrix whose structure allows for efficient solution of LWE problem. Given an LWE sample $s^{T} A+e^{T}$ and a type 2 trapdoor $R$ for $A$ :

$$
\left(s^{T} A+e^{T}\right) R=s^{T} G+\left(e^{T} R\right)
$$

This is an LWE sample for $G$ with error $e^{\prime}=e^{T} R$. This means we can use the same secret for constructing LWE samples for $G$ from the samples for $A$. Peikert \cite[Section 5.4.1]{peikert2016decade} states that ``\textit{gadget trapdoors are technically simpler to work with in SIS/LWE-based applications, computationally more efficient, and at least as powerful}'' as short basis trapdoors that we defined in the previous definition of `\textit{Type 1}' trapdoors.

\section{Signatures}

In the construction of lattice-based cryptographic signatures, there are two prominent families \cite{wang2023lattice} of signature schemes: the \textit{Hash-and-Sign} family and the \textit{Fiat-Shamir} family. They have distinct methodologies and offer different advantages in terms of security, efficiency, and implementation.

\subsection{Hash-and-Sign}

The hash-and-sign lattice-based signatures use a lattice trapdoor as the secret key. Goldreich Goldwasser and Halevi gave what is known as the GGH signature scheme \cite{goldreich1997public}. This is one of the earliest lattice-based signature proposals which uses hash-and-sign paradigm. With a good basis as the secret key, it computes the corresponding public key which is a bad basis. We refer to \cite[Section 4.2]{wang2023lattice} which describes the hash-and-sign scheme at a very high level to give the intuition:

- \textbf{Key Generation}: Generate a lattice $\mathcal{A} \subseteq \mathbb{Z}^{n}$ with a trapdoor $\mathrm{T}$ and a public representation P of $\mathcal{A}$. Return the public key $P$ and the secret key $T$.

- \textbf{Signing}: Given message $m$, compute $c=$ hash $(m) \in \mathbb{Z}^{n}$. Compute $v \in \mathcal{A}$ close to $c$ using $T$. Return the signature $\mathbf{s}=\mathbf{v}-\mathbf{c}$.

- \textbf{Verification}: Given message $m$ and its signature $s$, compute $c=$ hash $(m)$. Accept if $\mathbf{s}$ is short and $\mathbf{s}+\mathbf{c} \in \mathcal{A}$, otherwise reject.

The modern designs follow the provably secure framework by Gentry, Peikert and Vaikuntanathan \cite{GPV08} called GPV, in which the signature distribution is some discrete Gaussian, independent of the trapdoor i.e. the distribution of the signatures does not reveal any information about the trapdoor function which was used to create them. This provides a formal proof of unforgeability in the ROM \cite{peikert2016decade}. One of the signature schemes Falcon \cite{prest2020falcon}, which is to be standardized by NIST, is based on the GPV framework and works over NTRU \cite{hoffstein2003ntrusign} lattices.  It is fast and has the smallest signature size among other signature schemes. However it comes at the cost of complex construction where it is also non-trivial to understand and implement this scheme \cite{falcon_ppt}.

\subsection{Fiat-Shamir}

The Fiat-Shamir signature paradigm in lattices was given by Lyubashevsky in \cite{lyu08}. . At a high-level, the Fiat-Shamir approach transforms an interactive identification scheme into a non-interactive signature scheme. In lattices, it consists generating a non-interactive zero-knowledge (NIZK) proof  that the signer knows the secret short vector. To ensure that the signature distribution  leaks information  with negligible probability, Fiat-Shamir signatures uses rejection sampling. A very high-level description of Fiat-Shamir signatures is stated in \cite[Section 4.2]{wang2023lattice} as follows. Let $\chi$ be some distribution of small elements, and the algorithms:

- \textbf{Key Generation}: Generate $A \leftarrow U\left(\mathbb{Z}_{q}^{n \times m}\right)$ and $S \in \mathbb{Z}^{m \times k}$ of small coefficients. Return the public key $(A,T=AS)$ and the secret key $S$.

- \textbf{Signing}: Given message $m$, sample $y \leftarrow \chi^{m}$, compute $d=(A y \bmod q)$ and $c=$ hash $(d, m)$ where the hash domain is a set of short vectors. $\operatorname{Return}(z=S c+y, c)$ with certain probability, otherwise restart.

- \textbf{Verification}: Given message $m$ and its signature $(z, c)$, compute $d=(A z-T c \bmod q)$. Accept if $c=\operatorname{hash}(d, m)$ and $z$ is short, otherwise reject.

Dilithium \cite{ducas2018crystalsdilithium} is based on Module-LWE and uses the Fiat-Shamir approach for its design. Compared to Falcon, it has larger key and signature sizes but much simpler implementation \cite{huntcontextual}. Dilithium is selected as the primary signature algorithm for NIST PQC standardization. We now refer to \cite[Section 3.2]{huntcontextual} who gave a simplified description of the scheme.

\textbf{Key Generation}: A matrix $A$ and two short vectors $s_{1}$ and $s_{2}$ are obtained via random sampling. $A, s_{1}$, and $s_{2}$ are used to calculate $t=A s_{1}+s_{2}$. The public key $pk$ comprises of ($A$,$t$) and the secret key $sk$ is ($A, t, s_{1}$,$s_{2}$).

$$
\begin{gathered}
A \leftarrow R^{m \times n} \\
s_{1} \leftarrow S_{m}^{n}, s_{2} \leftarrow S_{m}^{m} \\
t=A s_{1}+s_{2} \\
p k=(A, t), s k=\left(A, t, s_{1}, s_{2}\right)
\end{gathered}
$$

\textbf{Signing}: The signing phase starts with choosing a short vector $y$, then computing $w$ $=$ $A y$. It uses the Fiat-Shamir transform to create the challenge polynomial $c$. The high bits of $w$ are taken with a message $M$, then challenge $c$ is  $H$($\mathrm{High}(w), M$) where $H$ is a hash function . For $z$ calculated as the sum $y+c s_{1}$, the scheme checks to see if $z$ or $w-c s_{2}$ reveals any secret information. This is the rejection sampling step and if there is any leakage, the signing process starts again. If not, we get the signature  as $z$ and $c$.

$$
\begin{gathered}
y \leftarrow S_{\gamma}^{n} \\
w=A y \\
c=\mathrm{H}(\mathrm{High}(w), M) \in B_{60} \\
z=y+c s_{1} \\
\|z\|_{\infty}>\gamma-\beta \vee\left\|\operatorname{Low}\left(w-c s_{2}\right)\right\|_{\infty}>\gamma-\beta \Longrightarrow \text { restart } \\
\operatorname{sig}=(z, c)
\end{gathered}
$$

\textbf{Verification}: Given public key $pk$ = $(A, t)$ and the signature $sig=(z,$ $c)$, a verifier takes the high bits of $Az-ct$ and computes a new challenge polynomial $c^{\prime}$ Here the rejection sampling step is done again to check $c^{\prime}$ against $c$ included in the received signature, to see if they match.

$$
\begin{gathered}
c^{\prime}=\mathrm{H}(\operatorname{High}(\underbrace{A z-c t}_{w-c s_{2}}), M) \\
\|z\|_{\infty} \leq \gamma-\beta \wedge c^{\prime}=c \Longrightarrow \text { accept }
\end{gathered}
$$

If it matches then $(z,$ $c)$ is indeed a signature on message $M$.

\cleardoublepage

\chapter{Lattice-based blind signatures}
\label{ch:chapter_5}

\section{State of the art}
\label{sec:5.1-soa}

The first lattice based blind signature was given by Ruckert \cite{ruckert2010lattice}. He followed the structure of three-move paradigm as seen in Schnorr or Okamoto-Schnorr blind signatures. This approach was extended and many publications followed but Hauck et al \cite{hauck2020lattice} showed that these had an error in their security proof. And these were not as safe as it was claimed earlier. He also gave the construction of the first provably secure lattice-based blind signature with the signature size of approximately 7.9 MB and communication cost of 34 MB. It only supported 7 signatures per verification key before it needed to be regenerated. The more recent works have tried to construct more efficient and provably secure blind signatures based on lattices where the number of signatures are not bounded.

Lyubashevsky et al. proposed a scheme which had signatures size of 150KB and communication cost of around 16 MB. \cite{lyubashevsky2022efficient}, thus making a huge improvement over previous works. Its security is based on the standard hardness of MSIS and MLWE assumptions but the signing cost increases with the number of signatures. This makes the scheme unusable in cases where we don't know the amount of signatures to be generated in advance. Agrawal et al. \cite{agrawal22} gave a very practical scheme which relies on the Gentry, Peikert and Vaikuntanathan (GPV) signature \cite{GPV08}. They build upon the Fischlin protocol \cite{fischlin2006round} which allows them to construct round-optimal signatures instead of three-moves. Round-optimal means that the signature will be created after two moves. This skips the possibility of concurrent executions in signing protocol and hence the resulting security issues. The signature size is around 45 KB and the scheme supports unbounded number of signatures. This efficient scheme was the result of using NIZKs by Lyubashevsky et al. \cite{lyubashevsky2022lattice} but its security is based on a new non-standard hardness assumption they introduced, which they call \textit{one-more-inhomogeneous} SIS assumption.

A new construction was proposed by del Pino and Katsumata \cite{del2022new}. By using the Fischlin construction \cite{fischlin2006round}, they were able to provide round-optimal signatures with size of 100 KB and communication cost of around 850 KB. The security of the scheme relies on the standard hardness of MSIS and MLWE assumptions while supporting unbounded number of signatures. The scheme can also be transformed into a partially-blind signature scheme, which allows the user and signer to append a common, agreed-upon message. This is useful for applications in e-cash and e-voting. The authors also gave a security proof in the Quantum-ROM (QROM) against quantum adversaries and they believe that by using the efficient techniques by Lyubashevsky et al. \cite{lyubashevsky2022efficient}, they can bring down the communication cost from 850 KB to around 100 KB. A shorter and efficient round-optimal blind signature scheme was proposed by Beullens et al. \cite{beullens2023lattice} with the signature size of just 22 KB. While following the Fischlin \cite{fischlin2006round} general idea, they modified the construction by Agrawal et al. \cite{agrawal22} to remove the new one-more ISIS assumption. It is also worth noting that they also use the techniques by Lyubashevsky et al. \cite{lyubashevsky2022efficient} to achieve this efficiency. 

Blind signature schemes can also be used to construct anonymous credentials. This mechanism allows users to prove the possession of necessary credentials required to use or access a service. Once these credentials are issued to the user, the user can choose to present them to the issuer for verification (by using zero-knowledge proof) instead of revealing all information. This allows for anonymity and selective-disclosure. Bootle et al. \cite{bootle2023framework} gave the construction of a blind signature scheme based on a newly proposed ISIS$_f$ assumption and showed how to turn the same protocol to an anoynmous credential scheme. The credential size was roughly 120 KB. The authors compare it to the recent works of \cite{jeudy2023lattice} and \cite{lai2023lattice} where the credential size was around 640KB and 500KB respectively. 

A new notion called non-interactive blind signature (NIBS) was recently put forward by Hanzlik \cite{hanzlik2023non} who made the observation that in many cases, the signed message was selected randomly. He explained that signer can create partial signatures called pre-signatures from the requester's public key. Later, the requester could extract a pair of blinded signature and the message from the pre-signature by using their private key. He also extended this idea to \textit{tagged} non-interactive blind signatures (T-NIBS) where a tag or unblinded metadata could be included with the signature. This functionality of T-NIBS resembles a partially-blind signature.

Taking the idea from the above, Baldimtsi et al. \cite{baldimtsi2024non} proposed the first practical construction of non-interactive blind signatures (NIBS) from lattices. Thus signer can issue pre-signatures without interacting with the requester, who later can extract valid blind signatures. They prove their security under a new variant of the one-more ISIS assumption \cite{agrawal22} called \textit{randomized} one-more ISIS (rOM-ISIS) with signature size of 68KB and communication cost of just 0.96 KB.  

\section{Round optimal lattice-based signature}
\label{sec:5.2}

The aim is to choose an appropriate scheme for the use case of e-cash. In a simple electronic cash or any other system using just blind signature, the signer has no control over the attributes except the ones already defined in the public key. It is ideal to use partially-blind signatures where the requester and signer can both agree on some common information to be added to blind signature. In the previous section, where the start-of-art was presented, one of the schemes made most sense to study. The construction by del Pino and Katsumata \cite{del2022new} meets the criteria. It supports unbounded polynomially many signatures and is provably secure. The scheme is based on standard assumptions as opposed to similar other works who either introduced new assumptions or doesn't include construction that provides partial blindness. Moreover, the authors established the security both in ROM and QROM, and even against malicious signer. Now we talk about the intuition, building blocks and construction of their scheme below:

\subsection{Intuition}

We saw that the Fischlin protocol \cite{fischlin2006round} is a nice starting point for many blind signature schemes and it provides round-optimal signature. This scheme by del Pino and Katsumata \cite{del2022new} also builds on same generic construction and supports unbounded many signatures.

In the presented blind signature construction, the signer or signing authority has a signing key $sk$, (which is kept secret) and a verification key $vk$ (which can be published).

To get a blind signature on a message $M$, the user or the requester submits a commitment com $\leftarrow \operatorname{Com}(\mathrm{M}$; rand) to the signer who gives the signature 
$\sigma \stackrel{\$}{\leftarrow} \operatorname{Sig}(\mathrm{sk}$, com $)$. 

The user also constructs a ciphertext ct $\leftarrow \operatorname{Enc}($ ek, com $\|$ rand $\| \sigma ;$ rand') using a PKE scheme and a NIZK proof $\pi$ that proves equations \ref{eq:com} and \ref{eq:Tandct} are equal. Here the statement is (vk, ek, ct, M) and the witness is (com, rand, $\sigma, $rand$^{\prime}$ ).

The user can now give $\Sigma=(\pi, \mathrm{ct})$ as the blind signature.

\begin{equation}
\label{eq:com}
\text { com }=\operatorname{Com}(\mathrm{M} ; \text { rand }) \wedge \operatorname{Verify}(\mathrm{vk}, \sigma, \text { com }) 
\end{equation}

\begin{equation}
\label{eq:Tandct}
    \top \wedge \mathrm{ct}=\operatorname{Enc}\left(\mathrm{ek}, \text { com } \| \text { rand } \| \sigma ; \text {rand}^{\prime}\right)
\end{equation}

The main restriction is giving an efficient lattice-based NIZK $\pi$ that proves equations \ref{eq:com} and \ref{eq:Tandct} are equal. It proves the possession of a valid signature on the hidden message ($com$ in this case).

The present lattice-based signatures doesn't have signing algorithm which are compatible with efficient and lattice-based NIZKs. To solve this limitation, the authors have pointed towards the technique by Pino et al. \cite{del2018lattice} in which they gave construction of efficient group signatures. 

\subsection{Building blocks}

A group signature allows a user to sign anonymously on a message, whose identity can be revealed later by the group manager in case of dispute. For the group user with identity $I \in [N]$ , the group manager generates a signature $\sigma \stackrel{\$}{\leftarrow} \operatorname{Sig}(\mathrm{sk}$, $I$$)$ which functions as the certificate for user $I$. When this user wants to sign on the behalf of the group, the user I generates $ct$ $\leftarrow \operatorname{Enc}($ ek, $I$; rand$^{\prime}$) using a PKE scheme and constructs an NIZK proof $\pi$ that proves

\begin{equation}
\label{eq:building}
\operatorname{Verify}(\mathrm{vk}, \sigma, I)=\top \wedge \mathrm{ct}=\operatorname{Enc}\left(\mathrm{ek}, I ; \text { rand }^{\prime}\right)    
\end{equation}

The statement is $X_{gs}$ is  $(\mathrm{vk}, \mathrm{ek}, \mathrm{ct})$ and the witness $W_{gs}$ is $\left(\sigma, I\right.$, rand$\left.^{\prime}\right)$. We can notice that though Eq. \ref{eq:building} is simpler than Eq. \ref{eq:com} and \ref{eq:Tandct}, it still provides a valid signature on a hidden message, which in this case is the identity $I$ of the user. 

In del Pino et al. \cite{del2018lattice}, the authors have used Boyen lattice signature \cite{boyen2010lattice} as the building block for their group signature scheme. In order to generate a credential for user $I$, the group owner or manager takes $I$ as the message and samples a short vector $\mathbf{e}$ satisfying

\begin{equation}
\label{eqcom}
\left[\mathbf{a}_{1} \mid \mathbf{a}_{2}+I \cdot \mathbf{g}\right] \mathbf{e}^{\top}=u 
\end{equation}

where $\mathbf{g}$ is the gadget matrix. The verification key consists of a random element $u \in R_{q}$ and vectors $\left(\mathbf{a}_{1}, \mathbf{a}_{2}\right) \in R_{q}^{k} \times R_{q}^{k}$, where $R_{q}$ is the polynomial ring $\mathbb{Z}_{q}[X] /\left(X^{d}+1\right)$. The signing key sk is a short basis $\mathbf{T}_{\mathbf{a}_{1}} \in R^{k \times k}$ that satisfies the relation $\mathbf{a}_{1} \mathbf{T}_{\mathbf{a}_{1}}=\mathbf{0} \bmod q$. 

In the case when the identity $I$ is pubic, the user can use lattice-based NIZKs based on standard assumptions of Module-SIS or Module-LWE to prove possession of $\mathbf{e}$, which is the certificate for that user. The proof is of the form $\overline{\mathbf{a}} \overline{\mathbf{e}}^{\top}=\bar{u}$, where the statement is $(\overline{\mathbf{a}}, \bar{u})$ and $\overline{\mathbf{e}}$ is the witness.

But in a group signature, the identity of the user $I$ is required to be hidden. Generating a lattice-based NIZK becomes hard as compared to the previous case (when $I$ is made public) because eq. (\ref{eqcom}) becomes a quadratic relation.

To tackle this limitation authors have used BDLOP commitment by Baum et al. \cite{baum2018more}. It is of the form com $=\left[\begin{array}{l}\mathbf{t}_{0} \\ \mathbf{t}_{1}\end{array}\right]=\left[\begin{array}{l}\mathbf{b}_{0} \\ \mathbf{b}_{1}\end{array}\right] \mathbf{R}+\left[\begin{array}{c}0 \\ I \cdot \mathbf{g}\end{array}\right]$, where $\mathbf{b}_{0}, \mathbf{b}_{1}$ is the commitment key, $\mathbf{R}$ is the commitment randomness, and $I \cdot \mathrm{g}$ is the message. 

From this com, we can use the lower part of $com$, i.e. $\mathbf{t}_{1}$ for eq. (\ref{eqcom}) and expand it as 
\begin{equation}
    {\left[\mathbf{a}_{1} \mid \mathbf{a}_{2}+I \cdot \mathbf{g}\right] \mathbf{e}^{\top} } =\left[\mathbf{a}_{1} \mid \mathbf{a}_{2}+\mathbf{b}_{1} \mathbf{R}+I \cdot \mathbf{g}\right] \mathbf{e}^{\top}-\mathbf{b}_{1} \mathbf{R} \mathbf{e}_{2}^{\top}
\end{equation}

and then finally express the left-hand side of Eq. (\ref{eqcom}) as 

\begin{equation}
\label{eq:final-rel}
{\left[\mathbf{a}_{1} \mid \mathbf{a}_{2}+I \cdot \mathbf{g}\right] \mathbf{e}^{\top} } =\left[\mathbf{a}_{1}\left|\mathbf{a}_{2}+\mathbf{t}_{1}\right| \mathbf{b}_{1}\right]\left[\begin{array}{c}
\mathbf{e}^{\top} \\
-\mathbf{R} \mathbf{e}_{2}^{\top}
\end{array}\right]
\end{equation}

We see that the term $\left[\mathbf{a}_{1}\left|\mathbf{a}_{2}+\mathbf{t}_{1}\right| \mathbf{b}_{1}\right]$ consists only of public elements. Hence, eq. (\ref{eqcom}) is shown as an M-SIS relation where $\left[\mathbf{a}_{1}\left|\mathbf{a}_{2}+\mathbf{t}_{1}\right| \mathbf{b}_{1}\right]$ is the statement and the witness vector is $\left[\mathbf{e} \mid-\mathbf{e}_{2} \mathbf{R}^{\top}\right]$. 

\textbf{Group signature by Pino et al.}:  The core idea is, given the message $msg$, the Boyen signature on $msg$ is converted to a signature on a commitment of $msg$. In the group signature setting, the message $msg$ = identity $I$ of the user is signed by the group manager and the user later can later show possession of valid signature (while also hiding the identity) by using a commitment. More concisely, the user transforms eq. (\ref{eqcom}) into eq. (\ref{eq:final-rel}), constructs an efficient NIZK proof $\pi$ for eq. (\ref{eq:final-rel}), and finally outputs the group signature $\Sigma=\left(\pi\right.$, com). 

\textbf{Using this technique for blind signatures}: The intuition behind the construction by del Pino and Katsumata \cite{del2022new} is that blind signature has opposite functionality as compared to the scheme in \cite{del2018lattice}. The signer will sign the message blindly by using commitment $com$ of the message and the user unblinds it to prove that he does possess a valid signature. Precisely the user constructs the BDLOP commitment com for the message $I$ which is sent to the signing authority. The signer pulls out $\mathbf{t}_{1}$ included in the commitment and signs it using the Boyen signature scheme. 

The signer samples a short vector $\mathbf{e}$ satisfying

$$
\left[\mathbf{a}_{1} \mid \mathbf{a}_{2}+\mathbf{t}_{1}\right] \mathbf{e}^{\top}=u
$$

The user then reverses the transformation in Eq. (\ref{eq:final-rel}) to obtain
\begin{equation}
\label{eq:blind-sig}
\left[\mathbf{a}_{1} \mid \mathbf{a}_{2}+\mathbf{t}_{1}\right] \mathbf{e}^{\top}=\left[\mathbf{a}_{1} \mid \mathbf{a}_{2}+\mathbf{b}_{1} \mathbf{R}+I \cdot \mathbf{g}\right] \mathbf{e}^{\top}=\left[\mathbf{a}_{1}\left|\mathbf{a}_{2}+I \cdot \mathbf{g}\right| \mathbf{b}_{1}\right]\left[\begin{array}{c}
\mathbf{e}^{\top} \\
\mathbf{R e}_{2}^{\top}
\end{array}\right]
\end{equation}

We can notice that in the above eq. (\ref{eq:blind-sig}), a public vector is multiplied by a short-secret vector. With the statement as $\left[\mathbf{a}_{1}\left|\mathbf{a}_{2}+I \cdot \mathbf{g}\right| \mathbf{b}_{1}\right]$ and the witness vector of $\left[\mathbf{e} \mid \mathbf{e}_{2} \mathbf{R}^{\top}\right]$, the signature can be a NIZK proof for Module-SIS relation.

\subsection*{One-more unforgeability}

First we have a brief overview of how Pino et al gave the proof of unforgeability of the group signature they proposed. The reduction simulates the group manager by sampling $\mathbf{a}_{1} \stackrel{\$}{\leftarrow} R_{q}^{k}$ and taking $\mathbf{a}_{2}$ as $\mathbf{a}_{2}=\mathbf{a}_{1} \mathbf{R}^{*}-I^{*} \cdot \mathbf{g}$ where $I^{*}$ is the guess for the user for which the adversary gives the forgery for. When the adversary queries the certificate for some user $I \neq I^{*}$ the reduction can use standard techniques \cite{cash2012bonsai} to sample a short vector for $\left[\mathbf{a}_{1} \mid \mathbf{a}_{2}+I \cdot \mathbf{g}\right]=\left[\mathbf{a}_{1} \mid \mathbf{a}_{1} \mathbf{R}^{*}+\left(I-I^{*}\right) \cdot \mathbf{g}\right]$.

If the adversary gives a forgery, consisting of proof $\pi$ and commitment $t_1$, the witness also satisfies  $\mathbf{t}_{1}=\mathbf{b}_{1} \mathbf{R}^{\prime}+I^{\prime} \cdot \mathbf{g}$, which is a valid BDLOP commitment. This reduction can break the MSIS problem if the adversary succeeds in unforgeability.

In the group signature, the message $I$ can be chosen by adversary. But it is not the case in the blind signature where the reduction should be able to sample from $\left[\mathbf{a}_1 \mid \mathbf{a}_2+\mathbf{t}_1\right]=\left[\mathbf{a}_1 \mid \mathbf{a}_1 \mathbf{R}^*-I^* \cdot \mathbf{g}+\mathbf{t}_1\right]$ for an arbitrary $\mathbf{t}_1$ hence the reduction cannot rely on the same trapdoor sampling techniques and presents a difficulty in simulating the real-world signer.

To tackle the same, the user also adds a NIZK proof of the well-formedness of $com$, $\pi_{\text {com }}$ along with the commitment $com$. It implies that $\mathbf{t}_{1}=\mathbf{b}_{1} \mathbf{R}^{\prime}+I^{\prime} \cdot \mathbf{g}$ for some short $\mathbf{R}^{\prime}$ and $I^{\prime}$.

\begin{figure}[h]
    \centering
    \includegraphics[scale=0.37]{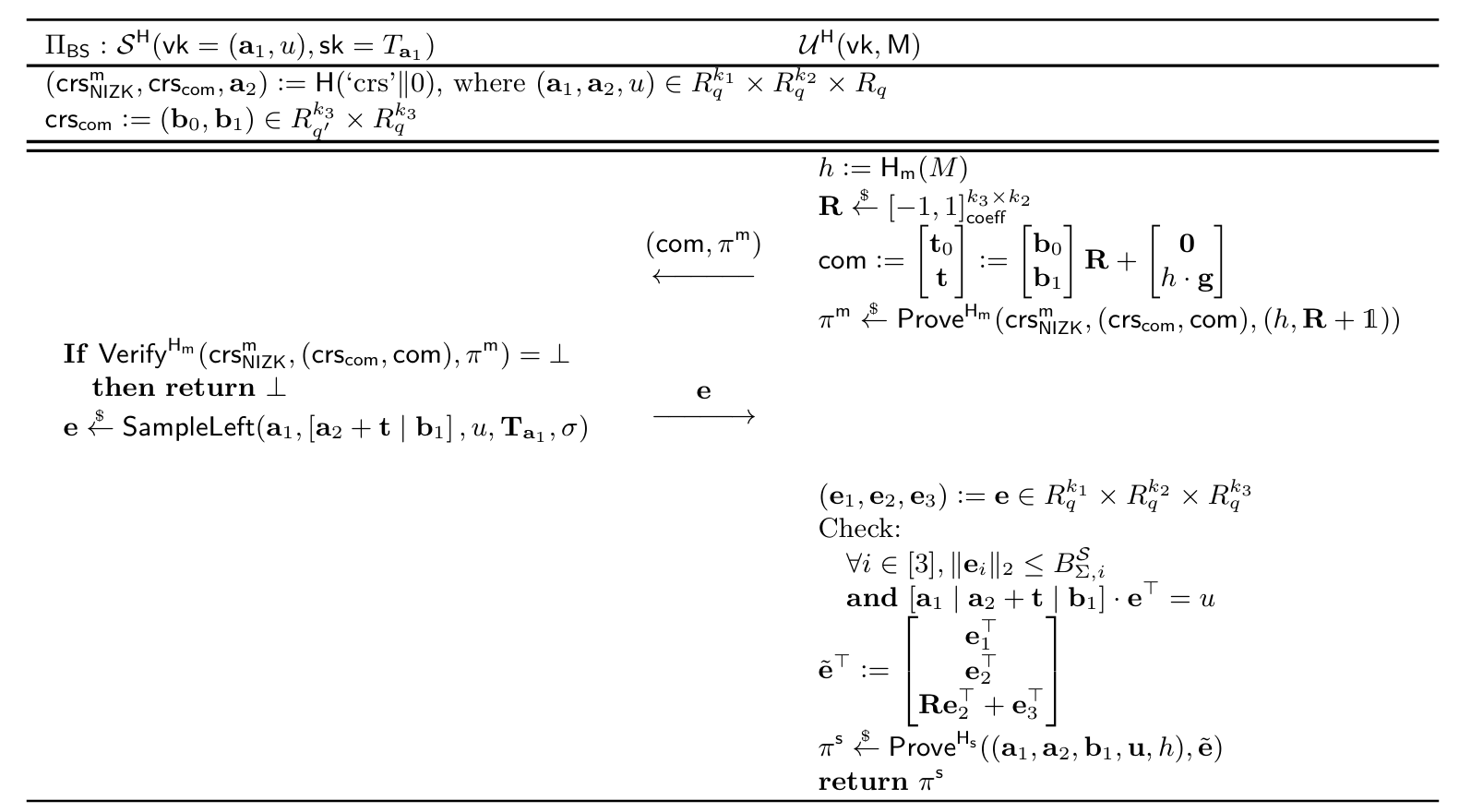}
    \caption{The blind signature scheme}
    \label{fig:BSS}
\end{figure}

The blind signature scheme can be summarized as follows.
The user constructs a (BDLOP) commitment $com$ for the message $M$ and adds a multi-proof straight-line extractable NIZK proof $\pi_{\text {com }}$ of its well-formedness. The signer receives $\left(\pi_{\text {com }}\right.$, com $)$ from the user and then samples a short vector $\mathbf{e}$ such that $\left[\mathbf{a}_{1}\left|\mathbf{a}_{2}+\mathbf{t}_{1}\right| \mathbf{b}_{1}\right] \mathbf{e}^{\top}=u$. Given e from the signer, the user transforms the signature verification equation into an MSIS relation following almost the same computation as in eq. (\ref{eq:blind-sig}), and outputs a standard NIZK proof $\pi$ for the MSIS relation as its signature. The full signature scheme is given in the Fig \ref{fig:BSS}

\subsection{Construction}
\label{ssec:cons}

Now we look at the different functions in the scheme:

\begin{figure}
    \centering
    \includegraphics[scale=0.37]{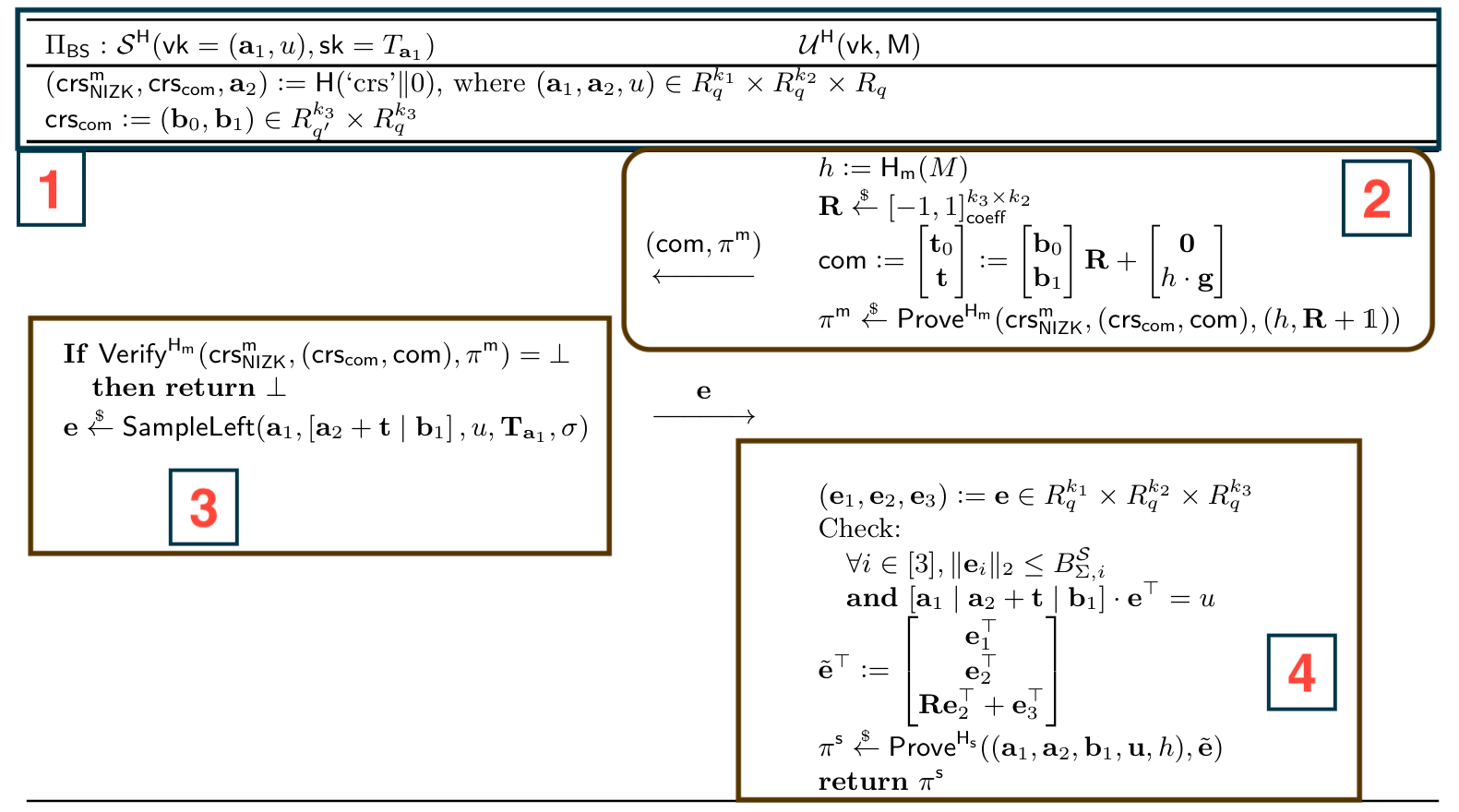}
    \caption{Components of the blind signature scheme}
    \label{fig:comp-bss}
\end{figure}

\textbf{Section 1: Setup} $\Pi_{\mathrm{BS}}$ is the blind signature scheme. The common random strings $\mathrm{crs}$ are derived correctly. With the trapdoor generation of $T_{\mathbf{a}_1}$, the pair of verification and secret key will be $\mathrm{vk}=(\mathbf{a}_1, u), \mathrm{sk}=T_{\mathbf{a}_1}$

\textbf{Section 2: Commitment and NIZK proof} For the message $M$ and a hash function as random oracle 
$\mathrm{H}_{\mathrm{m}}$, user gets $h:=\mathrm{H}_{\mathrm{m}}(M)$ for which they create a commitment $com$. Along with it, user constructs a NIZK proof $\pi^{\mathrm{m}}$ that proves the wellformedness of commitment $com$.

\textbf{Section 3: Blind signing} The verifier algorithm takes as input $crs$, a statement $X$ and a proof $\pi$. It either accepts or rejects. With a valid commitment $com$ and proof $\pi$, signer samples short vector $e$ such that $\left[\mathbf{a}_{1} \mid \mathbf{a}_{2}+\mathbf{t}_{1}\right] \mathbf{e}^{\top}=u$

\textbf{Section 4: Signature} The user creates a proof $\pi^{\mathrm{s}}$ that proves the knowledge of a short vector $\mathbf{e}$ which acts as the signature.

\subsection{Extension to partially blind signature}

Partially-blind signatures are an extension of blind signature schemes in which both the parties can agree to some common message which is included in signing. The blind signature scheme of del Pino and Katsumata et al. \cite{del2022new} can be modified to a partially blind signature scheme. 

All the building blocks are same as seen in Section \ref{ssec:cons} with minor modifications. The hash function $\mathrm{H}_{\mathrm{M}}$ is modified to take a message $\mathrm{M}$ and common message pair $\gamma$ as input $(\mathrm{M}, \gamma)$. A new hash function $\mathrm{H}_{\mathrm{M}_c}$ is also introduced.

In the signing phase, the signer samples short vector $e$ such that:

\begin{equation}
\left[\mathbf{a}_{1}\left|\mathbf{a}_{2}+\mathbf{t}\right| \mathbf{b}_{1}|\cdots| \mathbf{b}_{k_{2}}\right] \cdot \mathbf{e}^{\top}=\underline{u-\mathrm{H}_{\mathrm{M} c}(\gamma)}
\end{equation}

instead of $\left[\mathbf{a}_{1} \mid \mathbf{a}_{2}+\mathbf{t}_{1}\right] \mathbf{e}^{\top}=u$ earlier.

The user can construct a proof $\pi^{\mathrm{s}}$ as the signature $\Sigma$ which is binded to the common message $\gamma$. This partially-blind signature scheme satisfies partial blindness and one-more unforgeability \cite{del2022new}.

\section{Comparison to pre-quantum blind signatures}

We first note the signature and key size in 3072-bit RSA \cite{rfc9474} (which is the current recommended  parameter size by NSA) and 256-bit Elliptic Curve (EC) blind signature \cite{abe2000provably} which provides the same level of security. Both signature and keys are 384 bytes in size in RSA blind signatures while EC-based blind signatures have size of 128 bytes with key size of 32 bytes.

Recently, Katsumata et al. \cite{katsumata2023practical} presented two round-optimal blind signature schemes in the ROM based on group-based assumptions namely symmetric external Diffie-Hellman (SXDH), decisional Diffie-Hellman (DDH) and computational Diffie-Hellman (CDH) assumptions. In their first scheme, the signature and communication size is 447 bytes and 303 bytes respectively, thus having the smallest signature size among the prior schemes they have compared with. The second scheme achieves signature size of just 96 bytes while having communication size of 2.2KB. Both of these can be extended to partially-blind signatures. The efficiency in the first scheme was made possible by modifying the blind signature construction given by Fischlin \cite{fischlin2006round}. For their second scheme the authors improve upon the scheme by Blazy et al. \cite{blazy2013short} whose construction gave signatures of 96 bytes size but the communication cost increased linearly in message length.

Other comparable work includes the scheme by Hanzlik et al. \cite{hanzlik2023rai}. The authors presented a round-optimal blind signature scheme based on the CDH assumption that gave a signature size of 5 KB and communication size of 72 KB. Prior to this, Abe et al. \cite{abe2018improved} also showed an efficient scheme based on symmetric external Diffie-Hellman (SXDH) assumption with signature size of 5.5KB and around 1KB in communication cost.

\section{Application - electronic cash (eCash)}
\label{sec:5.4}

The first electronic money to have properties like physical cash was proposed by David Chaum \cite{chaumblind} whose scheme used a blind signature to achieve untraceability. In blind signature-based schemes, the customer could get the signature of the bank on the coin without disclosing any information about the coin, and spend it without revealing his identity to the merchant. But it is still desirable for the bank to include some information about the coin like validity, expiration date or a unique identifier. Using partially blind signature schemes, we can insert public information while keeping the private contents hidden. There is also the issue of $double$ $spending$. 

Unlike paper cash, electronic cash is inherently digital so it could easily be copied and reused. So, a malicious customer could spend it twice or more. Double spending can be checked by the bank in real-time, by keeping a database of the coin spent but it assumes that every participant is online. This comes at the expense of efficiency \cite{everaere2010double} where the bank is also a single point of failure and is also challenged to maintain optimal performance. Hence, achieving double spending detection and untraceability at the same time, is an issue in an off-line e-cash system \cite{baseri2013secure}. 

Everaere et al. \cite{everaere2010double} stated that, in a truly off-line setting, ``merchants are responsible for checking the validity of the coins on their own, which quite often can be difficult and costly''. The double spending might not get detected until the merchant goes online. Hence, the following question arises naturally:

\textit{What are the practical ways of achieving protection against double-spending while not online(preferably)?}

\subsection{P2P system for tackling double spending}
\label{sec:p2p_schemes}

We discuss two studies which rely on a peer-to-peer (P2P) design to detect or avoid double spending. The first one by Osipkov et al. \cite{osipkov2007combating} and the other one by Everaere et al. \cite{everaere2010double} 

Osipkov et al. \cite{osipkov2007combating} introduced a peer-to-peer system architecture to prevent double-spending. It doesn't require an on-line trusted party or tamper-resistant software or hardware. They mention that since merchants are the beneficiaries of a fair and fraud-less payment system, they can help in avoiding double spending. The idea lies on using merchant as a $witness$ to a transaction and they state that ``\textit{during creation, each coin is assigned in a random fashion to one of the merchants, which will serve as a witness for the validity of that coin}''. Expiration dates are also added in coin as public information, using a partially blind signature scheme.

In the proposed e-cash system, the merchants leave a security deposit with the broker $\mathcal{B}$, an entity where users can buy coin and merchants can exchange coins for bank credit/cash. When a user $\mathcal{U}$ obtains a coin $C$, it is also assigned one of the merchants $\mathcal{M}_C$ as the witness. When the user wants to spend the coin $C$ and sends it to a merchant $\mathcal{M}$, this merchant has to contact the witness $\mathcal{M}_C$ for a signature on payment transcript for the coin $C$. The witness merchant $\mathcal{M}_C$ can refuse to sign if the coin was seen before. Even if the witness merchant signs a payment transcript twice, they cannot cash the coins because the broker will detect this. In this case, the witness $\mathcal{M}_C$ is punished and their security deposit is used to pay the merchant $\mathcal{M}$.

\begin{figure}[h]
    \centering
    \includegraphics[scale=0.83]{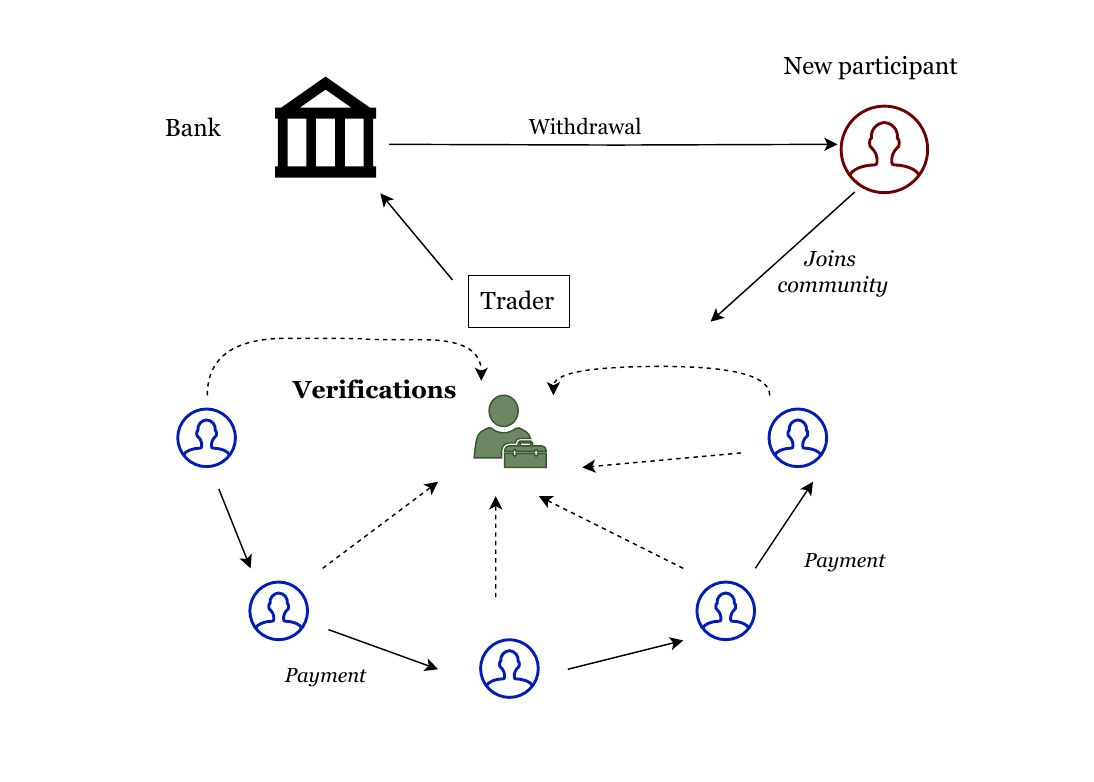}
    \caption{Scheme by Everaere et al.  \cite{everaere2010double}}
    \label{fig:p2p}
\end{figure}

The other study, Everaere et al.  \cite{everaere2010double} describes an approach which is based on risk management for protection against double spending. The goal of the scheme is to provide coverage to user in case of loss due to invalid coin as well as to minimize the communication with bank. This involves a $trader$, an entity who will take a fee per transaction but will also cover the potential risk of double spending.

They proposed the idea of communities which are managed by a trusted third-party, trader. These communities are autonomous groups where consumer and buyers can trade between themselves using the coins issued by the trader $T$. When a participant $\mathcal{P}$ (consumer or buyer) wants to join a community $\mathcal{C}$, they will exchange the bank certified coin $Coin_{bk}$ with trader certified coins $Coin_{t}$. The trader maintains a local database of the coins and these can be used for making payments within the community $\mathcal{C}$. A trader will also take some fee for providing coverage and insurance against double spending. When the participant $\mathcal{P}$ wants to exit this community $\mathcal{C}$ or enter a new community, they can redeem $Coin_{t}$ for bank certified $Coin_{bk}$. The novelty in their idea is that, they also address the risk taken by the trader where he may end up with double-spent coins from a malicious participant. A risk model is proposed which guarantees a minimal profit for the trader over a large number of transactions.

\subsection{Lattice-based eCash}

A good starting point for construction of lattice-based ecash system is the work by Bourse et al. \cite{bourse2019divisible}. The authors have proposed a generic framework for a $divisible$ ecash system using primitives like psuedo-random functions (PRF), digital signatures, commitment schemes and non-interactive zero-knowledge (NIZK) proofs. A divisible ecash system addresses the problem of paying exact amount. For example, in the case when a user has to pay \$2.5 for goods and he only has a coin worth \$5 or \$10, a merchant or bank will have to pay the rest back to user. On the other hand, if the user has divisible coins, i.e. coin of smallest denomination \$0.01, they can pay the exact amount and the transaction is therefore complete.

The authors then gave construction of a divisible/compact e-cash system. It is defined by the following algorithms and has three types of entities, the bank $\mathcal{B}$, a user $\mathcal{U}$ and a merchant $\mathcal{M}$.

The algorithms from their framework are described in a concise manner as follows:

- \textbf{Setup} $\left(1^\lambda, N\right)$ : On input a security parameter $\lambda$ and an integer $N$, this probabilistic algorithm outputs the public parameters $p p$ for divisible coins of global value $N$. The public parameters $p p$ are included in the other algorithms

- \textbf{BankKeygen()}: This probabilistic algorithm executed by the bank $\mathcal{B}$ outputs a key pair ($bsk$, $bpk$). It also sets $L$ as an empty list, that will store all deposited coins.

- \textbf{Keygen()}: This probabilistic algorithm executed by a user $\mathcal{U}$ as well as the merchant $\mathcal{M}$. It outputs a key pair for both user as ($usk$, $upk$) and for merchant as ($msk$, $mpk$).

- \textbf{Withdraw} $(\mathcal{B}$ ($bsk$, $upk$), $\mathcal{U}$ ($usk$, $bpk$)): This is an interactive protocol between the bank $\mathcal{B}$ and a user $\mathcal{U}$. At the end of this protocol, the user gets a divisible coin $C$ of value $N$ or outputs $\perp$ (in case of failure). The bank also stores the transcript of the protocol execution or outputs $\perp$.

- \textbf{Spend} $(\mathcal{U}($ $usk$, $C$, $bpk$, $V), \mathcal{M}(\mathrm{msk}$, $bpk$, $info$, $V))$ : This is an interactive protocol between a user $\mathcal{U}$ and a merchant $\mathcal{M}$. Here, $info$ denotes a set of public information associated to the transaction, by the merchant, and $V$ denotes the amount of this transaction. At the end of the protocol the merchant gets $Z$ (the coins for the product) along with a proof of validity $\Pi$ or outputs $\perp$.

- \textbf{Deposit} $(\mathcal{M}($ $msk$, $bpk$,($V$, $info$, $Z, \Pi)), \mathcal{B}($ $bsk$, $L$, $mpk$ $))$ : This is an interactive protocol between a merchant $\mathcal{M}$ and the bank $\mathcal{B}$.

There are two other algorithms which are used in the deposit protocol. The first one, $CheckDeposit$ checks if the deposit (the transcript and associated data) is valid. $Identify$ checks two transcripts and outputs the public key of the user in case double spending is detected.

To make the deposit, the merchant first sends a transcript $(V$, info, $Z, \Pi)$ along with some additional data $\mu$. The bank $\mathcal{B}$ first checks the validity of all these elements with the function $CheckDeposit$ and also that this merchant has not already deposited a transcript associated with $info$. This is done with the function $Identify$. If the data and transcripts are not valid, then $\mathcal{B}$ aborts and outputs $\perp$. If double spending is detected, then $\mathcal{B}$ returns another transcript $\left(V^{\prime}\right.$, info, $\left.Z^{\prime}, \Pi^{\prime}\right)$ along with the associated $\mu^{\prime}$. In this case, the bank can recover the serial numbers of the coin and compare it to the ones spent before. In case they were not double spent and are valid, then $\mathcal{B}$ adds it to the list $L$ (the list of deposited coins) and keeps a copy of ( $V$, info, mpk, $Z, \Pi$ ).

The above feature is the novelty of the framework given by \cite{bourse2019divisible} in which the authors use a new family of psuedo-random functions (PRFs) called $constrained$ PRFs. We refer the readers to the work \cite[Section 2]{bourse2019divisible} for more information. Taking ideas from this framework, Deo et al. \cite{deo2020lattice} then gave a construction for secure ecash system based on lattices. 

\subsection{Peer-to-peer system based eCash}

\begin{figure}[H]
    \centering
    \includegraphics[scale=0.86]{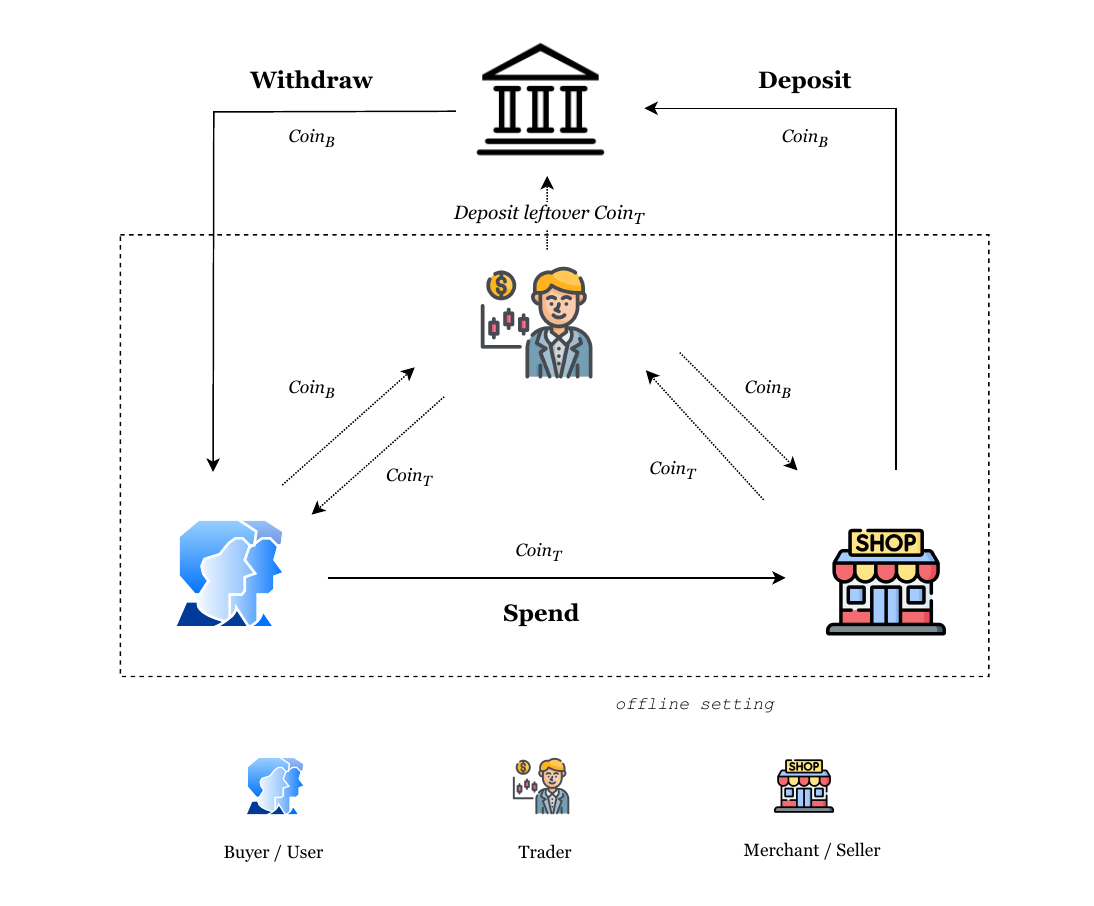}
    \caption{Peer-to-peer system of eCash}
    \label{fig:p2p-lattice-complete}
\end{figure}

From the two works discussed in the section \ref{sec:p2p_schemes}, the first work by  Osipkov et al. \cite{osipkov2007combating} proposes using merchants as \textit{witnesses} for the coins. But they don't talk about it in the context of divisible cash and the overall construction seems complicated. 

Witness assignment is one of the issues. The primary issue is, for spending an amount $V$, the $V$ serial numbers $\mathrm{SN}_{i_0}, \ldots, \mathrm{SN}_{i_{V-1}}{ }$ are derived. Then it is very much possible that different ranges in serial numbers might have different witnesses assigned to them. And each of them will be required to sign the respective transcript separately before any coins are spent. Hence, the work by Everaere et al. \cite{everaere2010double} deems to fit the framework of ecash better. Also, the main scheme including the protocols like \textit{Withdraw}, \textit{Spend} and \textit{Deposit} are already defined using lattice-based construction in the work by Deo et al. \cite{deo2020lattice}. Hence the peer-to-peer scheme of Everaere et al.  \cite{everaere2010double} can be used for the already constructed lattice-based ecash system. The figure \ref{fig:p2p-lattice-complete} shows our proposed peer-to-peer architecture for ecash where the entity $Trader$ is shown to be integrated with the scheme by Deo et al.\cite{deo2020lattice}. 

The user and merchants are part of a community where they can exchange the coins by bank denoted as $Coin_B$ with the coins of trader $Coin_T$. These coins will be used for making purchases. The exact steps for the spending protocol are defined in Everaere et al.  \cite{everaere2010double}. Whenever the user wants to spend coins to buy a product, the merchant generates a random value or a timestamp with the purchase order, we denote it as $Order_{Mer}$. This is sent to the user who then contacts the trader (with his coins and the $Order_{Mer}$) to get signature. The trader checks the validity of the coins, updates the local database for the coins spent and sends back the signature on the $Order_{Mer}$. This can be used be merchants to make purchases themselves or they can exchange for $Coin_B$ by interacting with the trader. 

\chapter{Conclusion}

We started with the concept of digital signatures and saw the construction of some classical signature schemes we use today. Then we moved on to blind signatures where we also described the models that are used to prove security of the schemes. Subsequently, we explained about structures called lattice and the main primitives which help us construct practical schemes that provide post-quantum security. In the last section, we show the state of art in lattice-based blind signatures and explained the construction of a round-optimal blind signature scheme in detail. Finally, we presented two peer-to-peer ecash systems that are present in the literature and saw that one of them might be a good fit to be constructed leveraging an already existing lattice-based ecash system. 

The objectives of this thesis are stated in the following, along with the sections that describe or try to answer them:

\textit{Give an overview of lattice-based cryptography and the main construction
tools available.}

\begin{itemize}
    \item The background of lattice-based cryptography and its primitives are described in Chapter \ref{ch:chapter_4}.
\end{itemize}

\textit{Review the recent lattice-based blind signature schemes and the new security
assumptions they are based on.}

\begin{itemize}
    \item In section \ref{sec:5.1-soa}, we have a detailed review of state-of-the-art in lattice-based blind signature schemes where signature sizes are shared along with the new security assumptions used.
\end{itemize}

\textit{Study some recent attacks and list the schemes affected by these attacks.}

\begin{itemize}
    \item One of the major attacks was the ROS attack. This is explained in the section \ref{sec:ros} along with the schemes which were affected by the same.
\end{itemize}

\textit{Describe the construction of the first blind signature scheme to have a formal
security proof in the quantum random oracle model.}

\begin{itemize}
    \item The scheme was proposed in the paper \cite{del2022new} by del Pino and Katsumata. We have attempted to explain the scheme in a detailed manner in the section \ref{sec:5.2}.
\end{itemize}

\textit{Review the approaches that tackle the problem of double spending in electronic cash and briefly evaluate their feasibility if lattice-based blind signatures are used.}

\begin{itemize}
    \item In the section \ref{sec:5.4}, we talk about ecash and see two models proposed to detect double spending. Finally we describe one lattice-based ecash scheme and discuss how the same scheme can be extended to include a peer-to-peer system that tackles double spending.
\end{itemize}

In this work, we proposed a design for ecash system which could benefit from the post-quantum security provided by lattice-based cryptography. The future work includes calculating the exact cost of realizing this model. This requires to calculate the cost incurred due to communication with the third party called trader. The P2P models were proposed more than a decade ago and seem practical in real-life scenario but we couldn't see any subsequent work based on them. Hence it is interesting to see if ecash systems can also make use of new and efficient lattice-based primitives \cite{lyubashevsky2022lattice}, \cite{lyubashevsky2021shorter}, while providing the desirable properties of real-physical cash.
\cleardoublepage

\cleardoublepage

\phantomsection
\addcontentsline{toc}{chapter}{\biblabel}
\printbibliography[title=\biblabel]
\cleardoublepage



\end{document}